\documentclass[pageno]{jpaper}
\pdfoutput=1

\newcommand{\asplossubmissionnumber}{45}
\newcommand\shade{\cellcolor{gray!20}} 

\usepackage[normalem]{ulem}

\usepackage{braket}
\usepackage{amsmath, amsfonts}
\usepackage{amsthm}
\usepackage{blkarray}
\usepackage{tikz}
\usepackage{svg}
\usepackage[table]{xcolor}
\usepackage[lined,boxruled]{algorithm2e}
\usepackage{qcircuit}
\usepackage{blkarray}
\usepackage{widetext}
\usepackage{authblk}

\newtheorem{theorem}{Theorem}

\begin{document}

\title{Minimizing State Preparations in Variational Quantum Eigensolver  \\ by Partitioning into Commuting Families}
\author[1]{Pranav Gokhale\thanks{Corresponding author: pranavgokhale@uchicago.edu}}
\author[2]{Olivia Angiuli}
\author[1]{Yongshan Ding}
\author[3]{Kaiwen Gui}
\author[4, 5]{Teague Tomesh}
\author[3, 4]{Martin Suchara}
\author[5]{Margaret Martonosi}
\author[1]{Frederic T. Chong}
\affil[1]{Department of Computer Science, University of Chicago}
\affil[2]{Department of Statistics, University of California, Berkeley}
\affil[3]{Pritzker School of Molecular Engineering, University of Chicago}
\affil[4]{Argonne National Laboratory}
\affil[5]{Department of Computer Science, Princeton University}

\date{\today}

\maketitle

\thispagestyle{empty}

\begin{abstract}
Variational quantum eigensolver (VQE) is a promising algorithm suitable for near-term quantum machines. VQE aims to approximate the lowest eigenvalue of an exponentially sized matrix in polynomial time. It minimizes quantum resource requirements both by co-processing with a classical processor and by structuring computation into many subproblems. Each quantum subproblem involves a separate state preparation terminated by the measurement of one Pauli string. However, the number of such Pauli strings scales as $N^4$ for typical problems of interest---a daunting growth rate that poses a serious limitation for emerging applications such as quantum computational chemistry. We introduce a systematic technique for minimizing requisite state preparations by exploiting the simultaneous measurability of partitions of commuting Pauli strings. Our work encompasses algorithms for efficiently approximating a MIN-COMMUTING-PARTITION, as well as a synthesis tool for compiling simultaneous measurement circuits. For representative problems, we achieve 8-30x reductions in state preparations, with minimal overhead in measurement circuit cost. We demonstrate experimental validation of our techniques by estimating the ground state energy of deuteron on an IBM Q 20-qubit machine. We also investigate the underlying statistics of simultaneous measurement and devise an adaptive strategy for mitigating harmful covariance terms.
\end{abstract}
\section{Introduction}

The present Noisy Intermediate-Scale Quantum (NISQ) era \cite{preskill2018quantum} is distinguished by the advent of quantum computers comprising tens of qubits, with hundreds of qubits expected in the next five years. Although several thousand logical error-corrected qubits, backed by millions of device-level physical qubits, are needed to realize the originally-envisioned quantum applications such as factoring \cite{shor1999polynomial} and database search \cite{grover1996fast}, a new generation of \textit{variational} algorithms have been recently introduced to match the constraints of NISQ hardware.

Variational Quantum Eigensolver (VQE) \cite{peruzzo2014variational} is one such algorithm that is widely considered a top contender, if not the top contender, for demonstrating a useful quantum speedup. VQE is used to approximate the lowest eigenvalue of a matrix that is exponentially sized in the number of qubits. This is a very generic eigenvalue problem with a wide class of applications such as molecular ground state estimation \cite{peruzzo2014variational}; maximum 3-satisfiability, market split, traveling salesperson \cite{nannicini2019performance}; and maximum cut \cite{moll2018quantum}. In this paper, we focus on the molecular ground state estimation problem which has already been demonstrated experimentally, though we underscore that the full range of VQE applications is very broad.

VQE solves a similar problem as Quantum Phase Estimation (QPE) \cite{kitaev1995quantum, cleve1998quantum}, an older algorithm that requires large gate counts and long qubit coherence times that are untenable for near-term quantum computers. VQE mitigates these quantum resource requirements by shifting some computational burden to a classical co-processor. As a result, VQE achieves low gate count circuits and error resilience, but at the cost of requiring many iterations where each iteration measures one of $O(N^4)$ terms.

This is a daunting scaling factor that poses practical limitations. It was observed that this $N^4$ scaling could be partly mitigated by performing simultaneous measurement: when two terms correspond to commuting observables, they can be measured in a single state preparation. Our work starts from this observation and we seek to exploit this idea to minimize the total number of state preparations needed.

Our specific contributions include:
\begin{enumerate}
    \item Efficient approximation algorithms for partitioning the $N^4$ terms into commuting families, i.e. approximating the MIN-COMMUTING-PARTITION.
    \item A circuit synthesis tool for simultaneous measurement.
    \item Statistical analysis of simultaneous measurement and a procedure for guarding against harmful covariance terms.
    \item Validation of these techniques through benchmarks, simulations, and experiments.
\end{enumerate}

The rest of this paper is structured as follows. Section~\ref{sec:background} presents relevant background material and Section~\ref{sec:prior_work} surveys prior work. Section~\ref{sec:analysis_of_commutativity} analyzes the commutativity of the terms of interest (Pauli strings) and Section~\ref{sec:min_clique_cover} presents a technique for minimizing the number of state preparations by mapping MIN-COMMUTING-PARTITION to a MIN-CLIQUE-COVER instance that can be approximated. Section~\ref{sec:linear_time_partitioning} develops an alternate technique that takes advantage of molecular Hamiltonian structure in order to approximate the MIN-COMMUTING-PARTITION with minimal classical overhead.

Section~\ref{sec:circuits_for_simultaneous_measurement} shows and analyzes the circuit synthesis procedure that allows simultaneous measurements between commuting Pauli strings. Section~\ref{sec:benchmark_results} presents results for our techniques on benchmark molecules and Section~\ref{sec:experimental_results} demonstrates experimental validation. Section~\ref{sec:covariance_reduction} studies the underlying statistics and discusses a strategy for detecting and correcting course if a partition is harmed by covariance terms. We make concluding remarks and propose future work in Section~\ref{sec:conclusion}.
\section{Background} \label{sec:background}
We assume an introductory-level knowledge of quantum computing. We refer newer readers to one of many excellent resources such as \cite{metodi2011quantum}, \cite{matuschak2019quantum}, or \cite{nielsen2010quantum}.

\subsection{Quantum Measurement}
\begin{figure}
    \centering
    \includegraphics[width=0.45\textwidth]{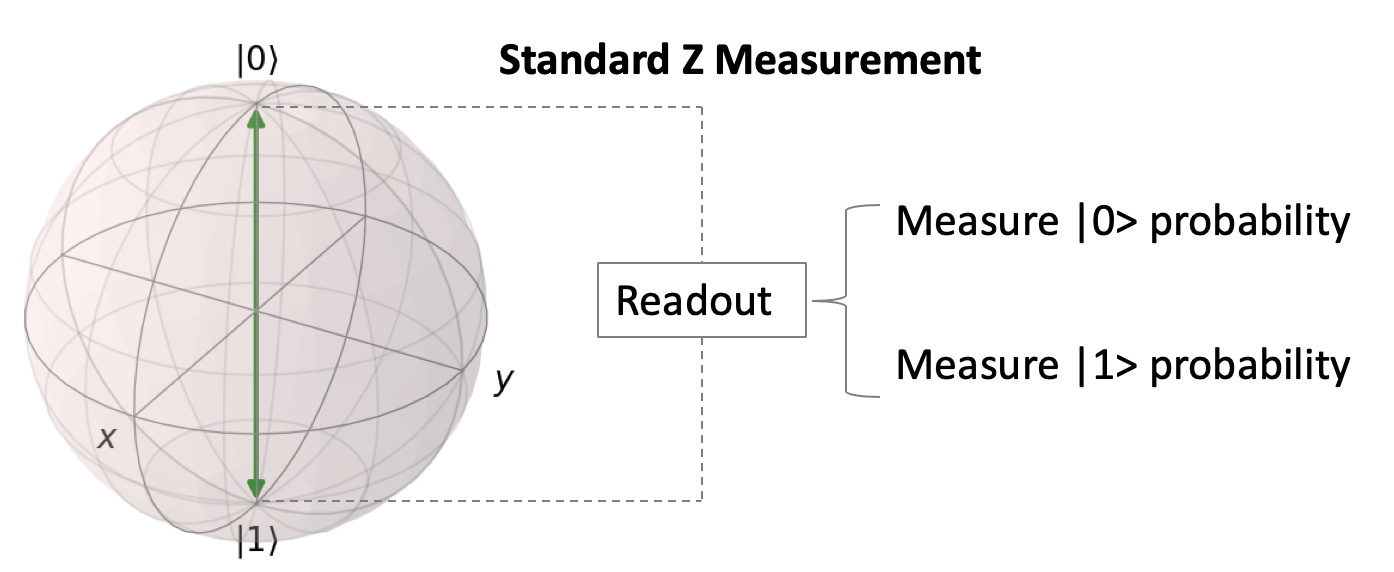}
    \caption{$Z$-basis (computational basis) measurement of a qubit yields $\ket{0}$ or $\ket{1}$ with a probability corresponding to the latitude of the qubit on the Bloch sphere.}
    \label{fig:standard_z_measurement}
\end{figure}

A standard procedure in quantum algorithms is to measure a qubit. In hardware, the standard measurement that can be performed is a measurement in the $Z$-basis, or computational basis. Figure~\ref{fig:standard_z_measurement} depicts such a measurement. The qubit's state is a point on the surface of the Bloch sphere---states with northern latitudes are close to the $\ket{0}$ state and southern latitudes are close to the $\ket{1}$ state. Measurement, or readout, causes the qubit to collapse to either the $\ket{0}$ or $\ket{1}$ state, with a probability dependent on the latitude.

\begin{figure}
    \centering
    \includegraphics[width=0.45\textwidth]{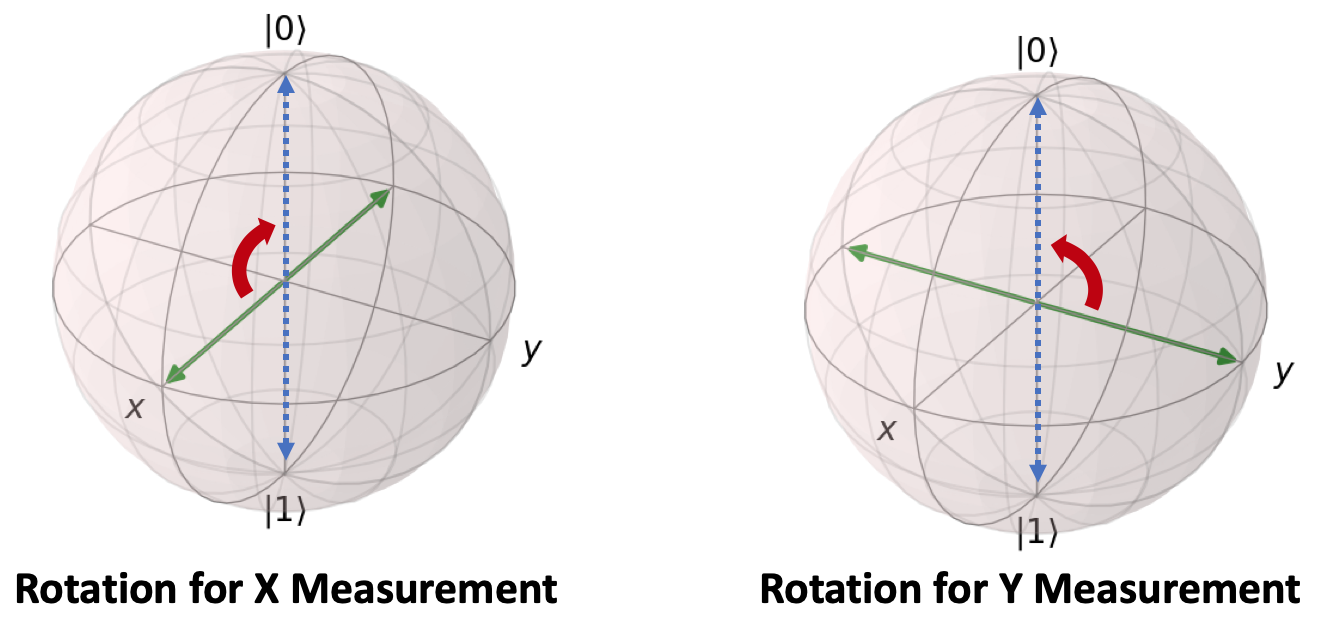}
    \caption{Measurement of the $X$ or $Y$ Pauli matrices requires us to first apply a unitary rotation operation that rotates the $X$ or $Y$ axis to align with the $Z$ axis. Subsequently, a standard $Z$-basis measurement yields the outcome of the $X$ or $Y$ measurement.}
    \label{fig:x_and_z_measurements}
\end{figure}

At a more mathematical level, the deeper meaning of measuring a qubit in the
$$Z=\begin{pmatrix}1 & 0 \\ 0 & -1 \end{pmatrix}$$
basis is to project the qubit's state onto the eigenvectors of the $Z$ operator, which are $\ket{0}$ and $\ket{1}$. In the same sense, we can measure other observables, such as the other two \textit{Pauli matrices}:
$$X = \begin{pmatrix} 0 & 1 \\ 1 & 0 \end{pmatrix} \quad \text{and} \quad Y = \begin{pmatrix} 0 & -i \\ i & 0 \end{pmatrix}$$
The eigenvectors of $X$ are termed $\ket{+}$ and $\ket{-}$, and they are antipodal points along $X$-axis of the Bloch sphere. Similarly, $Y$'s eigenvectors, $\ket{i}$ and $\ket{-i}$, are antipodal along the $Y$-axis. Since hardware cannot directly measure along these axes, measurements of $X$ ($Y$) are performed by first rotating the Bloch sphere with a unitary matrix so that the $X$ ($Y$) -axis becomes aligned with the $Z$-axis. These rotations are depicted in Figure~\ref{fig:x_and_z_measurements}. Subsequently, a standard $Z$-basis measurement can be performed, whose outcome can then be mapped to an effective $X$ ($Y$) measurement.

The specific rotation that accomplishes the $X$-to-$Z$ axis change is the $R_y(-\pi/2)$ transformation, which is typically captured in quantum circuits by the similar $H$ gate/matrix. The $Y$-to-$Z$ axis change is accomplished by the $R_x(\pi/2)$ transformation, which is typically captured \cite{pauli_measurements_2017} in quantum circuits by the $HS^{\dagger}$ gates/matrix.

The same general principle applies towards measuring observables across multiple qubits: measurement is accomplished by applying a quantum circuit that rotates the eigenvectors of the target observable onto the computational basis vectors. The unitary matrix for such a transformation is simply the one that has the orthonormal eigenvectors of the observable as column vectors. In our study, we will be interested in measuring Pauli strings, which are tensor products of Pauli matrices across multiple qubits.

\subsection{Simultaneous Measurement and Commutativity} \label{subsec:simultaneous_measurement}

From the preceding discussion, we can see that two observables can be measured simultaneously if they share a common eigenbasis, i.e. they are simultaneously diagonalizable. In this case, they can be measured simultaneously by applying the unitary transformation that rotates their shared eigenbasis onto the computational basis. In the case of Hermitian operators, such as the Pauli strings of interest to us, two observables share an eigenbasis if and only if they commute \cite[Chapter ~1]{shankar2012principles}, i.e. the order of their product is interchangeable.

Moreover, this relationship extends beyond simple pairs: given a family of pairwise commuting observables, there exists a shared eigenbasis that simultaneously diagonalizes \textit{all} of the observables (rather than it merely being a situation in which \textit{each} pair has a separate shared eigenbasis) \cite[Theorem 1.3.21]{horn2012matrix}.

In this paper, we will exploit this property to simultaneously measure multiple Pauli string observables with a single state preparation and measurement circuit. Notice that this problem is non-trivial because commutativity is not transitive (and hence, \textit{not} an equivalence class). Consequently, finding optimal partitions of commuting families is a hard problem, as we formalize later.

\subsection{Quantum Computational Chemistry}
Quantum computational chemistry has been a long targeted problem on the classical computer. Due to the limits of classical computing resources, we are only able to perform \textit{approximate} classical simulations. Examples include Hartree Fock ($O(N^{4})$ runtime \cite{HF_scaling}, only takes ground state orbitals into account), Density Functional Theory ($O(N^{3})$ runtime \cite{DFT_scaling}, but with even less precision), and Coupled Cluster Single-Double ($O(N^{6})+$ runtime \cite{CCSD_scaling}, only considers single and double excitations).

The way to achieve chemical accuracy is to use Full CI (full configuration interactions), which considers all necessary orbitals. Classically this will generally require $O(\binom{M}{N}) \rightarrow$ exponential runtime \cite{FCI_scaling}. On the other hand, quantum computation is able to encode an exponential amount of molecular information into a polynomial number of qubits and thereby achieve Full CI in polynomial time \cite{sugisaki2018quantum}.

\subsection{Variational Quantum Eigensolver (VQE)}

As mentioned previously, VQE can be applied to a wide class of problems that are solvable as minimum-eigenvalue estimation \cite{nannicini2019performance, moll2018quantum}. In this paper, we focus on the application that has received the most commercial and experimental interest: estimating molecular ground state energy. Within the molecular context, we use VQE to approximate the lowest eigenvalue of a matrix called the Hamiltonian that captures the molecule's energy configuration. The lowest eigenvalue is the ground state energy which has important implications in chemistry such as determining reaction rates \cite{eyring1935activated} and molecular geometry \cite{paulsen2004density}.

The Hamiltonian matrix for a molecule can be written in the second quantized fermionic form as \cite{mcardle2018quantum}

\begin{equation}
    H = \sum_{p=1}^N \sum_{q=1}^N h_{pq} a_p^{\dagger} a_q + \sum_{p=1}^N \sum_{q=1}^N \sum_{r=1}^N \sum_{s=1}^N h_{pqrs} a^{\dagger}_p a^{\dagger}_q a_r a_s
\label{eq:second_quantization}
\end{equation}

where $a^{\dagger}$ ($a$) is the fermionic raising (lowering) operator, and $N$ is the number of qubits and also the number of molecular basis wavefunctions considered. The $h_{pq}$ and $h_{pqrs}$ terms can be computed classically via electron integral formulas implemented by several software packages \cite{parrish2017psi4, sun2018pyscf, mcclean2017openfermion}. The second sum in Equation~\ref{eq:second_quantization} indicates that the fermionic form of the Hamiltonian has $O(N^4)$ terms \cite{wecker2014gate, hastings2014improving}. It can be translated to qubit form by an encoding such as Jordan-Wigner \cite{jordan1928pauli}, Parity \cite{seeley2012bravyi}, or Bravyi-Kitaev \cite{bravyi2002fermionic}, as we will discuss further in Section~\ref{sec:linear_time_partitioning}. The resulting qubit form will also have $O(N^4)$ terms, where each term is a Pauli string.

It is difficult to directly AND efficiently estimate $\braket{H}$, the expected energy of the Hamiltonian under an input state vector. The approach of VQE is to estimate it \textit{indirectly} but efficiently, by employing linearity of expectation to decompose $\braket{H}$ into a sum of $O(N^4)$ expectations of Pauli strings, which can each be computed efficiently. In the standard and original formulation of VQE, each of these Pauli strings is measured via a separate state preparation \cite{peruzzo2014variational}.

At its core, VQE can be described as a guess-check-repeat algorithm. Initially, the algorithm \textit{guesses} the minimum energy eigenvector of the Hamiltonian $H$. Then, it \textit{checks} the actual energy for the guessed eigenvector by summing expected values over the $O(N^4)$ directly measurable Pauli strings, as previously described. Finally, it \textit{repeats} by trying a new guess for the minimum energy eigenvector, with the assistance of a classical optimizer that guides the next guess based on past results. The potential quantum speedup in VQE arises from the fact that checking the energy on a classical computer would require matrix multiplication of an exponentially-sized state vector; by contrast, the energy can be estimated efficiently with a quantum computer by summing over the expected values of the $O(N^4)$ Pauli strings.

\begin{algorithm}
\SetAlgoLined
\KwResult{Approximate ground state energy, $\min_{\vec{\theta}} \braket{H}_{\psi(\vec{\theta})}$}
$\vec{\theta_1} \leftarrow$ random angles\;
$i \leftarrow 1$\;
\While{(not classical optimizer termination condition)}{
  \For{$j \in [O(N^4)]$}{
    \For{$O(1/\epsilon^2)$ repetitions}{
    Prepare $\psi(\vec{\theta_i})$\;
    Measure $\braket{H_j}_{\psi(\vec{\theta_i})}$\;
    }
  }
  $\braket{H}_{\psi({\vec{\theta_i}})} \leftarrow \sum_j \braket{H_j}_{\psi(\vec{\theta_i})}$\;
  Record $(\theta_i, \braket{H}_{\psi(\vec{\theta_i})})$\;
  $i$++\;
Pick new $\theta_i$ via classical optimizer\;

 }
 \caption{Variational Quantum Eigensolver (VQE)}
 \label{alg:vqe}
\end{algorithm}

Algorithm~\ref{alg:vqe} presents the pseudocode for VQE, under the standard `Naive' formulation where each Pauli string is measured separately. The resource complexity of VQE is clear from this code: the inner \texttt{for} loops run $O(N^4/\epsilon^2)$ times and each iteration requires a separate state preparation and measurement. The outer \texttt{while} loop termination condition is dependent on both the classical optimizer and the ansatz--we discuss the latter next.

\subsection{Unitary Coupled Cluster Single Double Ansatz}
Since the number of possible state vectors spans an exponentially large and continuous Hilbert space, we seek to restrict the family of candidate energy-minimizing states. Such a family is called an ansatz, and the ansatz state $\ket{\psi(\vec{\theta})}$ is parametrized by a vector of independent parameters, $\vec{\theta}$. Since VQE aims to run in polynomial time, the number of parameters should be polynomial. While our work in this paper is applicable to any ansatz, we focus our attention to the Unitary Coupled Cluster Single Double (UCCSD) ansatz, which has generally been the leading contender for molecular ground state estimation. In addition to having a sound theoretical backing (the coupled cluster approach is the gold standard for computational chemistry \cite{mcardle2018quantum, bartlett2007coupled}), UCCSD is more resilient to barren plateaus in the optimization landscape that are experienced by hardware-oriented ansatzes \cite{mcclean2018barren, mcardle2018quantum}. Recent work has also demonstrated the experimental superiority of UCCSD to other ansatz types \cite{mccaskey2019quantum}.

In terms of the number of qubits (which is also the number of molecular basis wavefunctions) $N$, the total gate count of UCCSD is $O(N^4)$ \cite{hempel2018quantum, lee2018generalized}, which can be parallelized in execution to $O(N^3)$ circuit depth. As a concrete scaling example, a recent 4-qubit, 2-electron UCCSD circuit construction required circuit depth of 100 gates, spanning 150 total gates \cite{mccaskey2019quantum}. This is already out of range of present machines---the experimental work thus far has required many symmetry reductions and approximations to implement UCCSD. The number of parameters in UCCSD, with respect to the number of electrons and wavefunctions is $O(N^2 \eta^2)$, or $O(N^4)$ under the standard assumption that these two terms are asymptotically related by a constant.

\subsection{Mutually Unbiased Bases} \label{subsec:mubs}

Finally, we give a brief overview of Mutually Unbiased Bases (MUB) \cite{schwinger1960unitary, klappenecker2003constructions}, a concept in quantum information theory that is connected to our overarching question of maximizing the information learned from a single measurement. In the case of qubits, MUBs describe a partitioning of the $4^N-1$ $N$-qubit Pauli strings (Identity is excluded) into commuting families of maximal size. For example, Table~\ref{tab:mubs} shows a MUB for the 2-qubit Pauli strings. Notice that each row corresponds to a commuting family. Also note that not all rows are created equal--in the first three rows, the shared eigenbasis features separable eigenvectors. In the last two rows, the shared eigenbasis has entanglement between the two qubits.

\begin{table}[h]
\begin{tabular}{cccc}
Operator 1 & Operator 2 & Operator 3 & Shared Eigenbasis \\ \hline \hline
ZZ         & IZ         & ZI         & Separable         \\ \hline
XX         & IX         & XI         & Separable         \\ \hline
YY         & IY         & YI         & Separable         \\ \hline
XY         & ZX         & YZ         & Entangled         \\ \hline
YX         & ZY         & XZ         & Entangled
\end{tabular}
\caption{MUB for two qubits. For the first 3 bases, the shared eigenbases has fully separable eigenvectors. The last 2 bases have fully entangled eigenvectors.}
\label{tab:mubs}
\end{table}

It is known that for $N$ qubits, there exists a MUB with $2^N+1$ rows and $2^N - 1$ Pauli strings per row. This is optimal in the sense that $2^N - 1$ is the maximum possible number of distinct Pauli strings (excluding Identity) within a commuting family. In Section~\ref{sec:min_clique_cover}, this result will give us insight into the bounds on our MIN-COMMUTING-PARTITION approach.
\section{Prior Work} \label{sec:prior_work}

Some of the theoretical aspects of our work were concurrently  and independently developed by two other research groups (our work was first presented a month earlier \cite{gokhale2019minimizing}). The four relevant papers, \cite{jena2019pauli} from Waterloo and \cite{verteletskyi2019measurement, yen2019measuring, izmaylov2019unitary} from Toronto all share with our work a high level goal of reducing the cost of VQE by exploiting the simultaneous measurability of commuting Pauli strings. In particular, \cite{jena2019pauli} maps the measurement cost reduction goal to a graph coloring problem. \cite{verteletskyi2019measurement} and \cite{yen2019measuring, izmaylov2019unitary}, which respectively consider Qubit-Wise Commutativity and General Commutativity (defined in Section~\ref{sec:analysis_of_commutativity}), treat measurement cost reduction as a minimum clique cover problem. The core ideas of these four papers can be compared to Sections \ref{sec:analysis_of_commutativity}-\ref{sec:min_clique_cover} and Appendix~\ref{app:np_hard} in this paper.

Our paper is differentiated by a systems perspective that gives explicit attention to the classical computation costs for compilation and transpilation, as well as quantum overheads. The graph algorithms discussed in \cite{jena2019pauli, verteletskyi2019measurement, yen2019measuring, izmaylov2019unitary} incur impractical classical costs that may undo potential speedups from simultaneous measurement. We remedy this issue by introducing problem-aware techniques that operate on molecular Hamiltonian graphs in linear time and hence preserve speedups, as discussed in Section~\ref{sec:linear_time_partitioning}. Also, in Section~\ref{sec:circuits_for_simultaneous_measurement}, we introduce a synthesis tool for simultaneous measurement circuits, in recognition of the fact that simultaneous measurement does incur a quantum overhead in additional gates and coherence requirements. To the best of our knowledge, this is the first synthesis tool that constructs simultaneous measurement circuits efficiently in both the classical compilation cost and in the quantum circuit complexity. Sections~\ref{sec:benchmark_results}  and \ref{sec:experimental_results} present benchmark results and experimental results validating that the classical and quantum costs of simultaneous measurement are worthwhile. Additionally, we study the statistics of simultaneous measurement in Section~\ref{sec:covariance_reduction} and demonstrate a constructive procedure to guard against corruption from covariance terms.

Prior to this month, strategies for simultaneous measurement in VQE had not been studied formally, aside from the initial suggestion of measurement partitioning in \cite{mcclean2016theory}. Most experimental implementations of VQE, for instance \cite{kandala2017hardware, nam2019ground, hempel2018quantum, kokail2018self}, did at least perform measurement partitioning on an ad hoc basis, via inspection of the Hamiltonian terms. Inspection is insufficient for larger molecules, because the underlying problem is NP-Hard, as described in Appendix~\ref{app:np_hard}. The improvement in these experimental works due to simultaneous measurement is indicated by the reduction from the \# Pauli Strings to QWC (Qubit-Wise Commutation) column in Table~\ref{tab:measurement_costs_past_vqe}. The last column considers General Commutation (GC) partitioning, which we introduce and evaluate in this paper. Even for the small molecules that have been studied experimentally thus far, GC achieves significant cost reductions over both Naive and QWC partitions.

\begin{table}[H]
\centering
\begin{tabular}{c|c|c|c}
Molecule             & \# Pauli Strings & QWC & GC \\ \hline \hline
H\textsubscript{2} \cite{kandala2017hardware} & 4                & 2   & 2 \\ \hline
LiH \cite{kandala2017hardware}                  & 99               & 25  & 9 \\ \hline
BeH\textsubscript{2} \cite{kandala2017hardware} & 164              & 44  & 8  \\ \hline
H\textsubscript{2} (Bravyi-Kitaev) \cite{hempel2018quantum} & 5               & 3 & 2 \\ \hline  H\textsubscript{2} (Jordan-Wigner) \cite{hempel2018quantum} & 14              & 5 & 2 \\ \hline
H\textsubscript{2}O \cite{nam2019ground}         & 21              & 3 & 3 \\ \hline
\end{tabular}
\caption{State preparation and measurement costs from prior VQE experiments that performed Pauli string partitioning on an ad hoc basis. \# Pauli Strings indicates the number of measurement partitions that would be needed naively. QWC expresses the number of Qubit-Wise Commuting partitions that were actually measured via ad hoc inspection---we propose a more formal partitioning procedure in Section~\ref{sec:min_clique_cover}. GC foreshadows the General Commuting partitions that our techniques described in Sections~\ref{subsec:gc} and \ref{sec:min_clique_cover}~-~\ref{sec:linear_time_partitioning} achieve.}
\label{tab:measurement_costs_past_vqe}
\end{table}

In software implementations, both the OpenFermion \cite{mcclean2017openfermion} and Rigetti PyQuil \cite{smith2016practical} libraries were recently augmented with functions for simultaneous measurement via Qubit-Wise Commutation: \texttt{group\_into\_tensor\_product\_basis\_sets()} and \texttt{group\_experiments()} respectively. However, these software implementations do not consider General Commutativity and suffer from at least $N^8$ scaling in runtime, which may undo the potential speedup from simultaneous measurement.

An alternative perspective on the reduction of measurement cost in VQE was introduced in \cite{bravyi2017tapering} which takes the approach of transforming molecular Hamiltonians to \textit{create} commutativity and reduce the number of qubits needed. Another prior paper \cite{izmaylov2019revising} operates in a related mathematical setting, using feedforward measurements to create QWC (though we note that feedforward measurements are equivalent to standard unitary transformations by the principle of deferred measurement \cite{nielsen2010quantum}).

Aside from state preparation and measurement costs, recent work has focused on improving other elements of the VQE pipeline. In the classical stage, \cite{nannicini2019performance, mcclean2016theory, moseley3bayesian} describe improvements to the classical optimizer and \cite{shi2019optimized, gokhale2019partial} present techniques for optimized pulse-level compilation. At the quantum stage, \cite{lee2018generalized, dallaire2018low} propose improvements to ansatzes and \cite{mcclean2016theory, otten2019accounting} demonstrate procedures for error mitigation. We note that all of these techniques apply to orthogonal stages of the VQE pipeline and therefore can compose directly on top of our work.
\section{Analysis of Commutativity} \label{sec:analysis_of_commutativity}

We analyze the commutativity of the terms present in Hamiltonian decompositions. Two terms $A$ and $B$ commute, if their \textit{commutator} is 0:
$$[A, B] := AB - BA = 0 \rightarrow AB = BA$$
As mentioned in Section~\ref{subsec:simultaneous_measurement}, two commuting terms are simultaneously diagonalizable by a shared eigenbasis.

In our case, the terms in an $N$-qubit Hamiltonian are Pauli strings, which are $N$-fold tensor products of the Pauli matrices,
$$
I = \begin{pmatrix} 1 & 0 \\ 0 & 1  \end{pmatrix}, \hspace{.23em}
X = \begin{pmatrix} 0 & 1 \\ 1 & 0 \end{pmatrix}, \hspace{.23em}
Y = \begin{pmatrix} 0 & -i \\ i & 0 \end{pmatrix}, \hspace{.23em}
Z = \begin{pmatrix} 1 & 0 \\ 0 & -1 \end{pmatrix}.
$$
Pauli strings are also referred to in other literature as members of the Pauli Group, $G_N$.

We seek to analyze when two Pauli strings commute. While most of these results are known, they are usually discussed in the context of the stabilizer formalism and quantum error correction. We present the elements relevant to VQE here, with foreshadowing of our key techniques.

\subsection{Single Qubit Case}
First, let's note the commutation relations for single qubit Pauli matrices:
\begin{itemize}
    \item $I$ commutes with everything else. Specifically, $[I, I] = [I, X] = [I, Y] = [I, Z] = 0$.
    \item $X$, $Y$, and $Z$ commute with themselves. $[X, X] = [Y, Y] = [Z, Z] = 0$.
    \item The other pairs form a cyclic ordering. In particular, $[X, Y] = iZ$, $[Y, Z] = iX$, $[Z, X] = iY$. Flipping the commutator bracket order negates the result.
\end{itemize}

\subsection{Qubit-Wise Commutativity (QWC)}
The simplest type of commutativity is Qubit-Wise Commutativity (QWC). Two Pauli strings QWCommute if at each index, the corresponding two Pauli matrices commute. For instance, $\{XX, IX, XI, II\}$ is a QWC partition, because for any pair of Pauli strings, both indices feature commuting Pauli matrices.

As mentioned in Section~\ref{sec:prior_work}, QWC has been leveraged in past experimental work for small molecules \cite{kandala2017hardware, nam2019ground, hempel2018quantum, kokail2018self} by ad hoc inspection of the Hamiltonian terms. However, Appendix~\ref{app:np_hard} demonstrates that optimally partitioning Pauli strings into QWC families is NP-Hard, so an efficient approximation algorithm is needed for larger Hamiltonians with more Pauli strings.

QWC is also referred to in other work as Tensor Product Basis (TPB) \cite{kandala2017hardware, mcclean2017openfermion, smith2016practical}, recognizing the fact that for a family of QWC Pauli strings, the vectors in the simultaneous eigenbasis can be expressed as a tensor product across each qubit index, with no entanglement. As shown in Section~\ref{sec:circuits_for_simultaneous_measurement}, this makes simultaneous measurement very easy for QWC partitions.


\subsection{General Commutativity (GC)} \label{subsec:gc}
QWC is sufficient but not necessary for commutation between Pauli strings. For example, $\{XX, YY, ZZ\}$ is a commuting family, even though none of the pairs are QWC---at both indices the Pauli matrices always fail to commute. The most general rule for commutation of two Pauli strings is that they must fail to commute at an \textit{even} number of indices---2 in the example of $\{XX, YY, ZZ\}$. We refer to this most general form of commutativity as General Commutativity (GC), and its proof is below. Note that QWC is simply the subset of GC corresponding to the case where the number of non-commuting indices is 0 (which is even).

\begin{theorem}
Consider two $N$-qubit Pauli strings, 
$$A = \bigotimes_{j=1}^N A_j \text{ and } B = \bigotimes_{j=1}^N B_j$$
where $A_j, B_j \in \{I, X, Y, Z\}$. $A$ and $B$ commute (GC) iff $A_j$ and $B_j$ fail to commute on an even number of indices.
\end{theorem}
 
\begin{proof}
For Pauli matrices that don't commute, $A_i B_i = - B_i A_i$.
Thus, we can write $AB$ as
$$AB = \bigotimes_{j=1}^N A_j B_j = \bigotimes_{j=1}^N \begin{cases}
B_j A_j & \text{if $[A_j, B_j] = 0$} \\ 
- B_j A_j & \text{if $[A_j, B_j] \neq 0$}
 \end{cases}
= (-1)^k BA$$
where $k$ is the number of indices where $[A_j, B_j] \neq 0$. For $AB$ to equal $BA$, we require $(-1)^k = 1$, which requires $k$ to be even. Thus, $A$ and $B$ commute iff $A_j$ and $B_j$ commute on an even number of indices.
\end{proof}

\begin{figure}[h]
    \centering
    \includegraphics[width=0.45\textwidth]{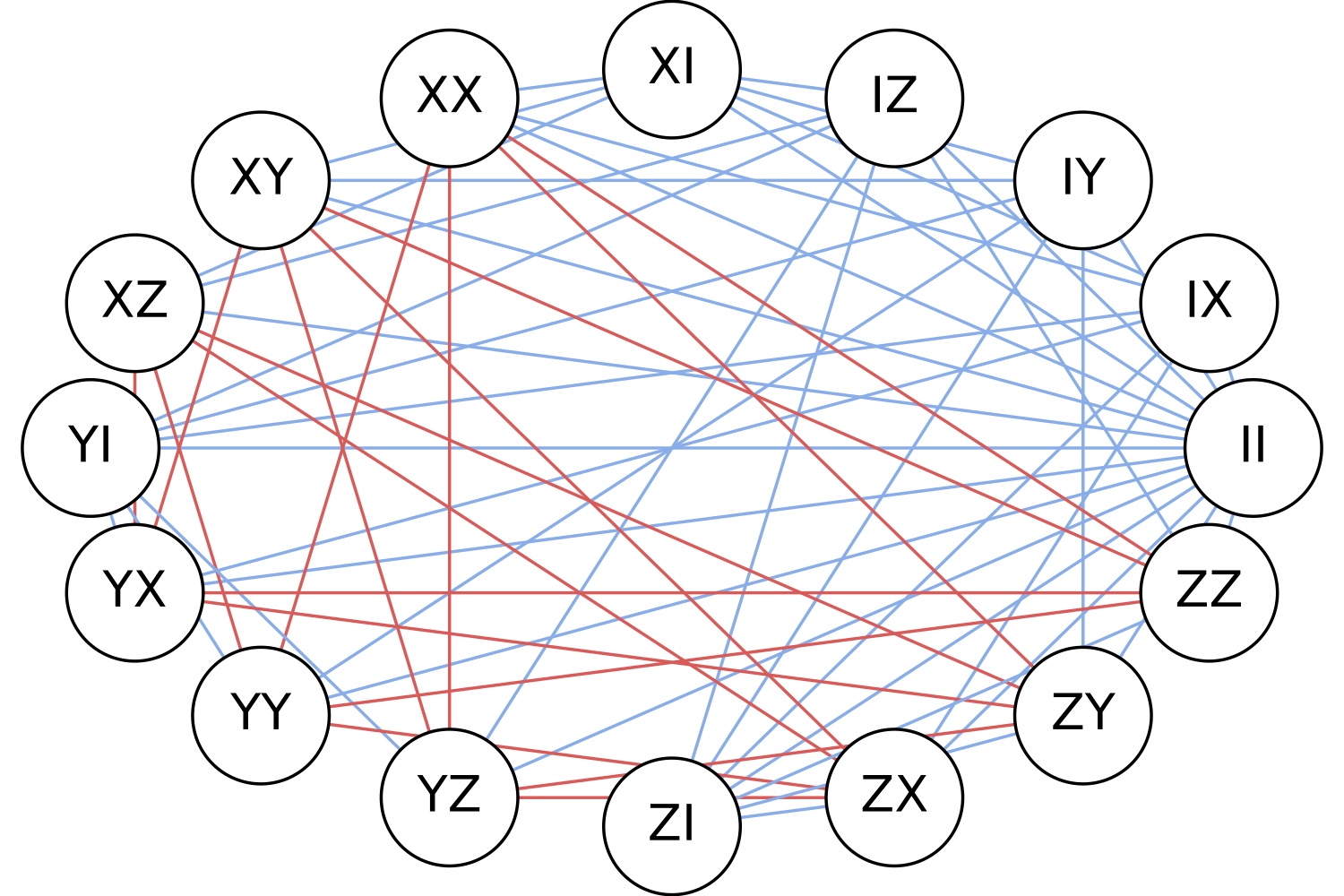}
    \caption{This is the commutation graph (also known as a compatibility graph \cite{kirby2019contextuality}) for all 16 2-qubit Pauli strings. An edge appears when two Pauli strings commute. The blue edges indicate Pauli strings that commute under QWC (which is a subset of GC). The red edges commute under GC-but-not-QWC.}
    \label{fig:two_qubit_pauli_graph}
\end{figure}

Figure~\ref{fig:two_qubit_pauli_graph} depicts the commutation relationships between all 16 2-qubit Pauli strings. Edges are drawn between Pauli strings that commute---a blue edge indicates that the pair is QWC and a red edge indicates that the pair is GC-but-not-QWC. The $II$ identity term QWCommutes with every other Pauli string.

\section{MIN-CLIQUE-COVER on Hamiltonian} \label{sec:min_clique_cover}
We refer to our core problem of interest as MIN-COMMUTING-PARTITION: given a set of Pauli strings from a Hamiltonian, we seek to partition the strings into commuting families such that the total number of partitions is minimized. While the underlying structure of Pauli matrices and their commutation relationships raises the possibility that MIN-COMMUTING-PARTITION may be efficiently solvable, it turns out to be NP-Hard, as we prove in Appendix~\ref{app:np_hard}. Moreover, MIN-COMMUTING-PARTITION is hard even when we only consider the restricted commutativity of QWC. Thus, the ad hoc QWC partitioning techniques from past experimental work \cite{kandala2017hardware, nam2019ground, hempel2018quantum, kokail2018self} are likely to have limited potential for larger molecules.

\begin{figure}[h!]
    \centering
\begin{tikzpicture}
\node[inner sep=0pt] (H2_graph) at (-4.25,0)
    {\includegraphics[width=.45\textwidth]{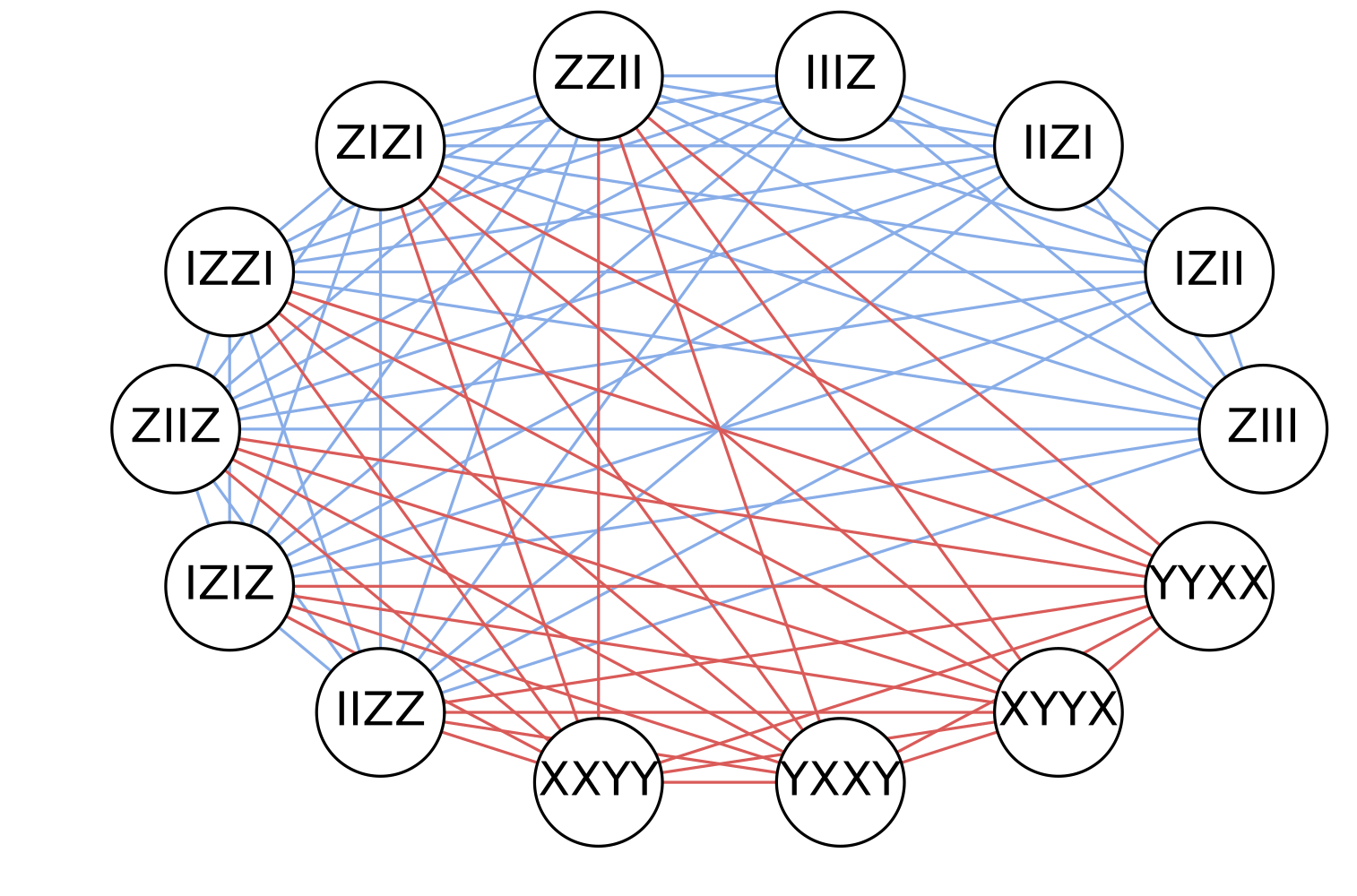}};
\node[inner sep=0pt] (H2_QWC_Min_Clique_Cover) at (-6.16,-5.5)
    {\includegraphics[width=.24\textwidth]{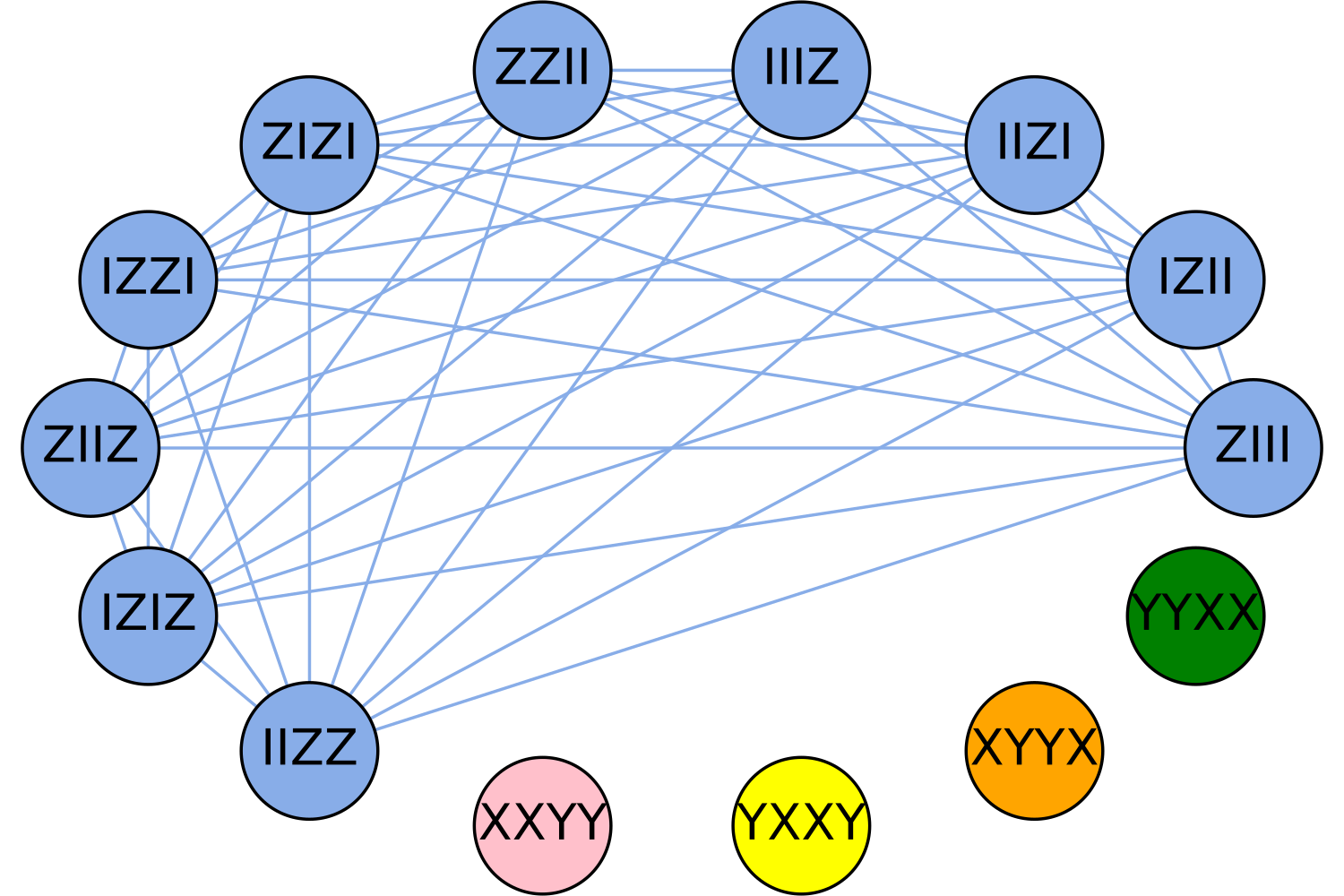}};
\node[inner sep=0pt] (H2_Full_Min_Clique_Cover) at (-1.84,-5.5)
    {\includegraphics[width=.24\textwidth]{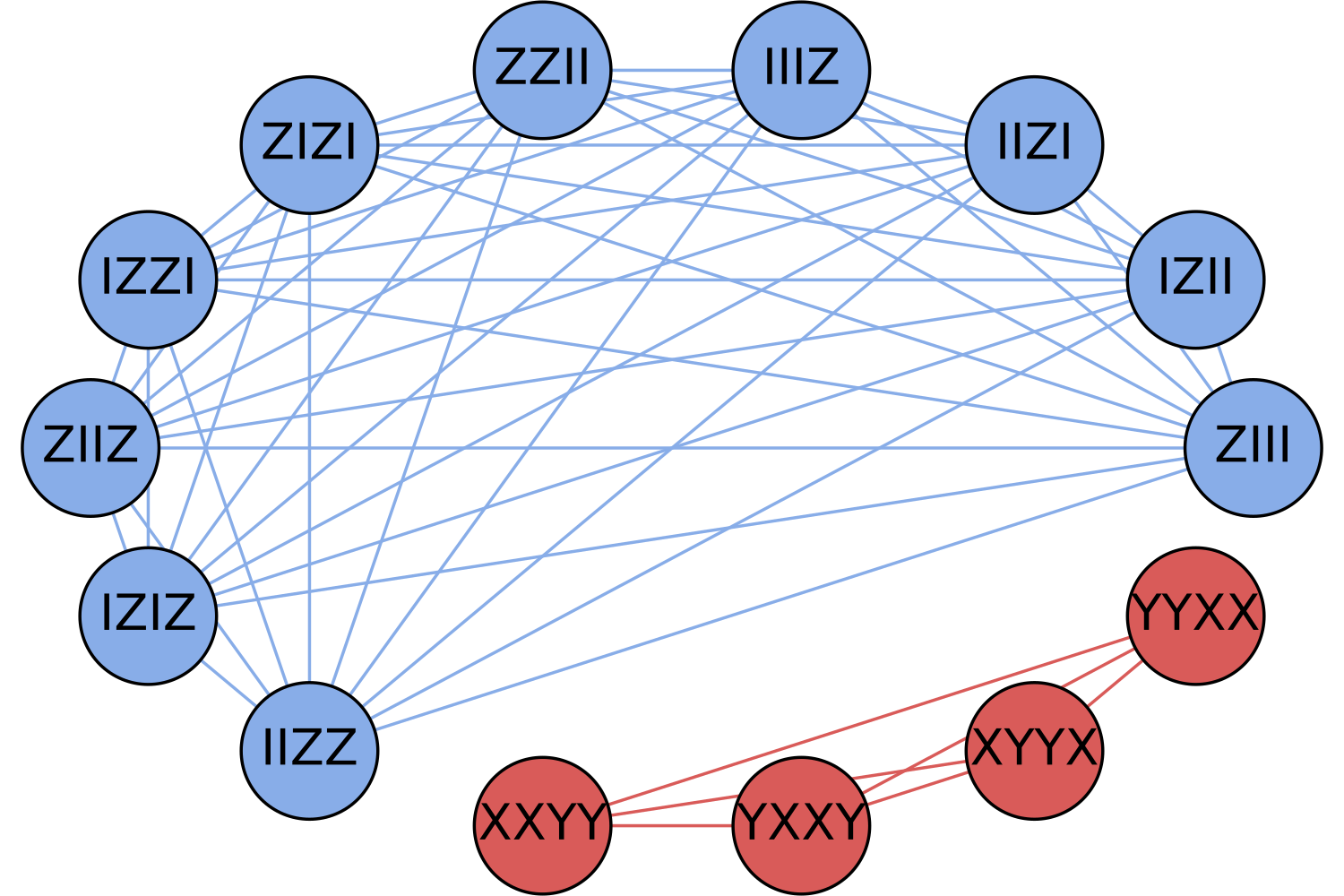}};
\draw[->,thick] (-4.15,-2.3) -- (H2_QWC_Min_Clique_Cover.north)
    node[midway,fill=white] {QWC};
\draw[->,thick] (-3.85,-2.3) -- (H2_Full_Min_Clique_Cover.north)
    node[midway,fill=white] {All edges (Full)};
\end{tikzpicture}
    \caption{The top commutation graph shows both QWC (blue) and GC-but-not-QWC Commuting (red) relationships between the Pauli string's in LiH's Hamiltonian. The vertex colors in the bottom two graphs indicate MIN-CLIQUE-COVERs using only QWC edges (left) or using all edges (right). The reduction in measurement partitions from Naive (measuring each Pauli string separately) to QWC to GC is $14\rightarrow5\rightarrow2$.}
    \label{fig:LiH_min_clique_cover}
\end{figure}

Instead of solving MIN-COMMUTING-PARTITION exactly, we approximately solve it by mapping to a graph problem as suggestively expressed by the graph representation in Figure~\ref{fig:two_qubit_pauli_graph}. Observe that cliques (fully connected subgraphs where each pair of Pauli strings commutes) are relevant because all of the strings in a clique can be measured simultaneously. Therefore, we seek the MIN-CLIQUE-COVER, i.e. the smallest possible set of cliques whose union spans all vertices. As an example, Figure~\ref{fig:LiH_min_clique_cover} shows the commutation graph for LiH's 4-qubit Hamiltonian and its MIN-CLIQUE-COVERs using QWC edges and using GC edges.

MIN-CLIQUE-COVER, in its decision version, is one of the classic Karp NP-Complete problems \cite{karp1972reducibility}, so efficiently finding the minimal possible clique cover for a general graph is unlikely. Moreover, finding a guaranteed ``good'' clique cover approximation is also NP-Hard for general graphs \cite{zuckerman2006linear}. However, molecular Hamiltonian graphs are highly structured owing both to features of the Pauli commutation graph \cite{planat2007pauli} and to patterns in the Pauli strings that arise in molecular Hamiltonians (we explicitly address and exploit the latter in Section~\ref{sec:linear_time_partitioning}).  This suggests that MIN-CLIQUE-COVER approximation algorithms may yield reasonably good results. Before discussing the approximation algorithms we used, we discuss bounds on the MIN-CLIQUE-COVER and the relationship to whether the partitions are QWC or GC.

\subsection{Bounds via MUBs}
Note that $2^N$ separate Pauli strings can be measured via a single simultaneous measurement. For instance, consider the $2^N$ set of Pauli strings of form $(I \text{ or } Z)^{\otimes N}$. All such Pauli strings can be simultaneously measured by simply measuring in the Z basis on each qubit. This example is suggestive of the power of simultaneous measurement. In the graph picture, it means that cliques exist of size $2^N$, which means that simultaneous measurement can lead to an exponential reduction in quantum cost relative to Naive separate measurements.

In the case of VQE, we will consider graphs that have only a polynomially sized ($O(N^4)$) number of Pauli strings. It is still enlightening to consider the MIN-CLIQUE-COVER on the $N$-qubit graph comprising all $4^N - 1$ possible Pauli strings (in this analysis, we exclude $I^{\otimes N}$ which commutes with everything else). Per the MUB formalism introduced in Section~\ref{subsec:mubs} and as suggested in the previous paragraph, a clique of Pauli strings can contain at most $2^N - 1$ vertices. This suggests that at least $2^N + 1$ cliques are needed to cover all $4^N - 1$ possible Pauli strings on $N$ qubits. In fact, this lower bound is exactly attainable---a MUB is exactly such a covering of all $N$-qubit Pauli strings by disjoint cliques. Again, this illustrates the potential of simultaneous measurement---a square root reduction is achieved in the total number of state preparations and measurements needed to cover all possible $N$-qubit Pauli strings.

Many of the partitions produced by MUBs have entanglement in the shared eigenbasis: for example, the bottom two rows of the MUB in Table~\ref{tab:mubs}. This means that the MIN-CLIQUE-COVER corresponding to a MUB requires GC edges and not just QWC edges. Next, we further discuss the advantage of GC over QWC.

\subsection{QWC vs. GC} \label{subsec:qwc_vs_gc}
GC captures a much denser commutation graph than QWC does, and therefore has more opportunities for larger cliques and thereby smaller clique covers.

We first consider the commutation graph of QWC, over all possible $N$-qubit Pauli strings; this graph has $4^N$ vertices. Given a Pauli string with $I$ on $k$ indices, it QWC commutes with exactly $4^k \cdot 2^{N-k} - 1 = 2^{N+k} - 1$ other Pauli strings: on the `partner' string, the $k$ indices are unrestricted and the $N-k$ indices can either match the original Pauli matrix or be $I$ (we subtract 1 to not count the original Pauli string). Since there are $\binom{N}{k} 3^{N-k}$ terms with $I$ on exactly $k$ indices, we see that
$$|E| = \sum_{k=0}^N \frac{\binom{N}{k} 3^{N-k} (2^{N+k} - 1)}{2} = \frac{10^N-4^N}{2}$$
This corresponds to an asymptotic graph density of
$$\lim_{N \to \infty} \frac{|E|}{|V|(|V|-1)/2} = \lim_{N \to \infty} \frac{(10^4-4^N)/2}{4^N(4^N-1)/2} = \lim_{N \to \infty} (5/8)^N = 0.$$

In other words, the QWC graph is extremely sparse. By contrast, the GC graph is dense: consider two random Pauli strings. The indicator variable denoting whether the two strings commute on the $i$th index is a Bernoulli random variable. Therefore, the GC commutation graph corresponds to when the sum over $N$ such independent variables is even, i.e. when a Binomial random variable is even. Asymptotically, this occurs with $\frac{1}{2}$ probability---thus the asymptotic graph density for GC is $\frac{1}{2}$, much denser than for QWC.

Although GC leads to smaller MIN-CLIQUE-COVERs than QWC, QWC does have cheaper simultaneous measurement circuits, as we will see in Section~\ref{sec:circuits_for_simultaneous_measurement}. However, the cost of GC simultaneous measurement will still turn out to be favorable, because circuit costs in VQE are dominated by the ansatz preparation.

\subsection{Approximation Algorithms Tested}
In our benchmarking, we performed MIN-CLIQUE-COVERs using the Boppana-Halldórsson algorithm \cite{boppana1992approximating} included in the NetworkX Python package \cite{hagberg2008exploring}, as well as the Bron-Kerbosch algorithm \cite{bron1973algorithm} which we implemented ourselves. These heuristics approximate a MAX-CLIQUE whose vertices are marked; we then recurse on the residual unmarked graph, repeating until all vertices are marked. We also used the \texttt{group\_into\_tensor\_product\_basis\_sets()} approximation implemented by OpenFermion \cite{mcclean2017openfermion}---this approximation is a non-graph-based randomized algorithm that only finds QWC partitions. Section~\ref{sec:benchmark_results} presents results across a range of molecules and Hamiltonian sizes.

While the benchmark results indicate promising performance in terms of finding large partitions, it is critical to also consider the classical computation cost of performing the MIN-CLIQUE-COVER approximation. First, the Bron-Kerbosch algorithm has a worst case exponential runtime. Therefore, its optimality should be interpreted as a soft upper bound on how well other standard approximation algorithms can approximate a MIN-CLIQUE-COVER. The Boppana-Halldórsson algorithm's runtime is polynomial but is not well studied. Our benchmarks and theoretical analysis indicate roughly quadratic scaling in graph size. Some polynomial benchmarks considered in the other concurrent work scale as much as cubically in the graph size.

However, this poses a problem---the Hamiltonian graph has $N^4$ terms, so a quadratic or cubic runtime in the number of vertices implies $N^8$ or $N^{12}$ scaling in classical precomputation time. Beyond simply implying impractical scaling rates, these runtime ranges may exceed the quantum invocation cost of VQE, in which case, we'd be better off just running VQE in the Naive fashion. In particular, recall that the UCCSD ansatz has $O(N^3)$ circuit depth after parallelization and that naively, $O(N^4)$ state preparations are needed per ansatz. The total quantum invocation cost of VQE therefore scales as $N^7$ multiplied by the number of ansatz states explored, though we note that both the ansatz exploration and the naive $O(N^4)$ measurements could be parallelized given multiple quantum machines. The number of ansatz states explored is an open question that depends on the classical optimizer, the ansatz type, and the variational landscape. Nonetheless, we can make rough estimates by noting that the VQE ansatz has $O(N^4)$ parameters, and rough theoretical results suggest anywhere from $O(N^4)$ iterations under the default SciPy optimization settings \cite{jones2001scipy} to $O(N^{12})$ under matrix inversion techniques. Further work is needed to understand the exact cost of VQE, but there is a strong case that standard graph approximation algorithms may have higher asymptotic cost than simply executing VQE naively without simultaneous measurement optimization. In the case of many expensive MIN-CLIQUE-COVER approximation algorithms, it seems likely that it would be better to simply skip the partitioning step and just measure the Pauli strings naively.

In the next section, we remedy this concern by presenting a MIN-COMMUTING-PARTITION approximation that exploits our knowledge of the structure of molecular Hamiltonians and their encodings into qubits. The resulting approximation algorithm runs in $O(N^4)$ time (linear in the number of Pauli strings, i.e. the graph size), which is safely below the quantum invocation cost of VQE.
\section{Linear-Time Partitioning} \label{sec:linear_time_partitioning}
As discussed in the previous section, standard MIN-CLIQUE-COVER approximations may be unsuitable since the classical cost of partitioning can exceed the quantum cost from naively running VQE. This motivates us to inspect features of molecular Hamiltonians and develop a new partitioning strategy accordingly. At a high level, our new strategy is context-aware and attacks the MIN-COMMUTING-PARTITION problem at a different abstraction level, namely the encoding stage from fermionic Hamiltonian to qubit Hamiltonian. By contrast, the previous approximations are unaware of molecular properties.

For convenience, we repeat Equation~\ref{eq:second_quantization} for the molecular Hamiltonians:
$$H = \sum_p^N \sum_q^N h_{pq}  a^{\dagger}_p  a_q +  \sum_p^N \sum_q^N \sum_r^N \sum_s^N h_{pqrs}  a^{\dagger}_p a^{\dagger}_q a_r a_s$$
where $a^{\dagger}$ and $a$ denote raising and lowering operators that act on fermionic modes.

The $N^4$ scaling of the number of terms in the Hamiltonian is clear from the second summation. In particular, the asymptotically-dominant terms are of form $a^{\dagger}_p a^{\dagger}_q a_r a_s$ with $p \neq q \neq r \neq s$. These $O(N^4)$ terms are known as the \textit{double excitation} operators \cite{whitfield2011simulation}. At the scale of smaller molecules, the $O(N)$ terms of form $a_p^{\dagger} a_p$ and the $O(N^2)$ terms of form $a_p^{\dagger} a_q^{\dagger} a_p a_q$ are frequent. These are termed the \textit{number} and \textit{number-excitation} operators respectively. We will treat both the asymptotically-dominant terms and the frequent-for-small-molecules terms in this section.

The commutation relationships of fermions are different from the commutation relationships of qubits. Thus, an encoding step is needed to convert the fermionic Hamiltonian into a qubit Hamiltonian. We consider the most common \cite{mcardle2018quantum} such encodings: Jordan-Wigner \cite{jordan1928pauli}, Parity \cite{seeley2012bravyi}, and Bravyi-Kitaev \cite{bravyi2002fermionic}.

\subsection{Jordan-Wigner}
Under the Jordan-Wigner encoding, we make the fermion-to-qubit transformations:
$$a_p \rightarrow \frac{X_p + i Y_p}{2} Z_{p-1} ... Z_0, \quad a_p^{\dagger} \rightarrow \frac{X_p - i Y_p}{2} Z_{p-1} ... Z_0 $$
with $I$ on every other index.

\subsubsection{Double excitation operators}. For the asymptotically dominant $O(N^4)$ terms of form $a^{\dagger}_p a^{\dagger}_q a_r a_s$ (WLOG, $p > q > r > s)$, we end up with the 16 Pauli strings matching the regular expression: $$(X_p|Y_p)Z_{p-1}...Z_{q+1}(X_q | Y_q) (X_r | Y_r)Z_{r-1}...Z_{s+1}(X_s | Y_s)$$

\begin{figure}[h]
    \centering
    \includegraphics[width=0.45\textwidth]{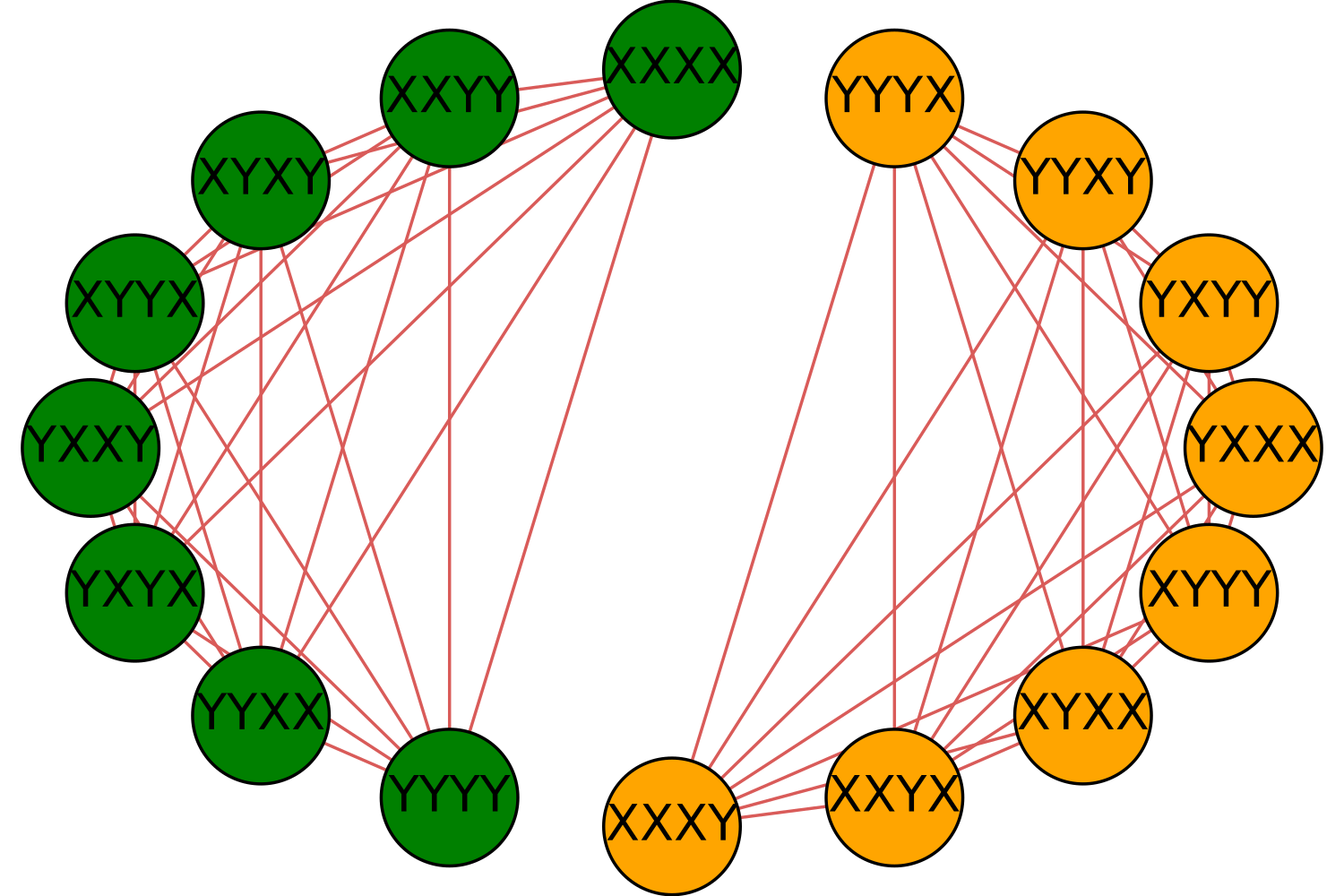}
    \caption{The 16 relevant Pauli strings in the Jordan-Wigner encoding of $a^{\dagger}_p a^{\dagger}_q a_r a_s$ have a MIN-CLIQUE-COVER of size 2.}
    \label{fig:jw_pqrs_clique}
\end{figure}

Thus, we see that the Jordan-Wigner transformation turns each of the $N^4$ fermionic terms into a sum over 16 Pauli strings. Moreover, these 16 Pauli strings are disjoint from the ones generated by a $a^{\dagger}_{p'} a^{\dagger}_{q'} a_{r'} a_{s'}$ term. Consider the commutation graph of the 16 Pauli strings. All indices except for $p, q, r,$ and $s$ immediately commute, so the commutativity graph only needs to consider the $p, q, r,$ and $s$ indices. Figure~\ref{fig:jw_pqrs_clique} depicts the commutation graph, which has a MIN-CLIQUE-COVER of 2. Thus, this yields a strategy for reducing the number of measurement partitions by 8x: we collect all Pauli strings from fermionic terms of form $a^{\dagger}_p a^{\dagger}_q a_r a_s$ (and from the 4! permutations of the indices) and measure them using 2 GC partitions instead of 16 Naive partitions.

For molecular Hamiltonians, we generally expect to have $h_{pqrs} = h_{srqp}$, because of the nature of these calculations via integrals and the fact that electrons are indistinguishable. In this case, only 8 terms arise (as noted in another context by \cite{whitfield2011simulation}), specifically the green 8-clique in Figure~\ref{fig:jw_pqrs_clique}. Thus again, we can achieve an 8x reduction.

\subsubsection{Number and number-excitation operators}

While the 8-fold reduction in the partitions of the $O(N^4)$ $pqrs$ terms is the asymptotic bottleneck, we also note a useful reduction for the smaller terms which are significant for smaller molecules.

For the $O(N)$ number operators of form $a_p^{\dagger} a_p$, multiplying out the Jordan-Wigner encoding yields the Pauli string $Z_p$. For the $O(N^2)$ number-excitation operators of form $a_p^{\dagger} a_q^{\dagger} a_p a_q$, the Jordan-Wigner encoding yields the Pauli string $Z_p Z_q$.

Observe that all of these Pauli strings commute and therefore can be simultaneously measured. Moreover, they are QWC, so the simultaneous measurements are cheap, as we will see in Section~\ref{sec:circuits_for_simultaneous_measurement}. While this result may appear obvious from inspection of small molecular Hamiltonians, which have many Pauli strings of form $I...IZI...I$, we underscore that it is not obvious to a context-unaware MIN-CLIQUE-COVER approximation.

\subsection{Parity Encoding}
\begin{figure}[h]
    \centering
    \includegraphics[width=0.45\textwidth]{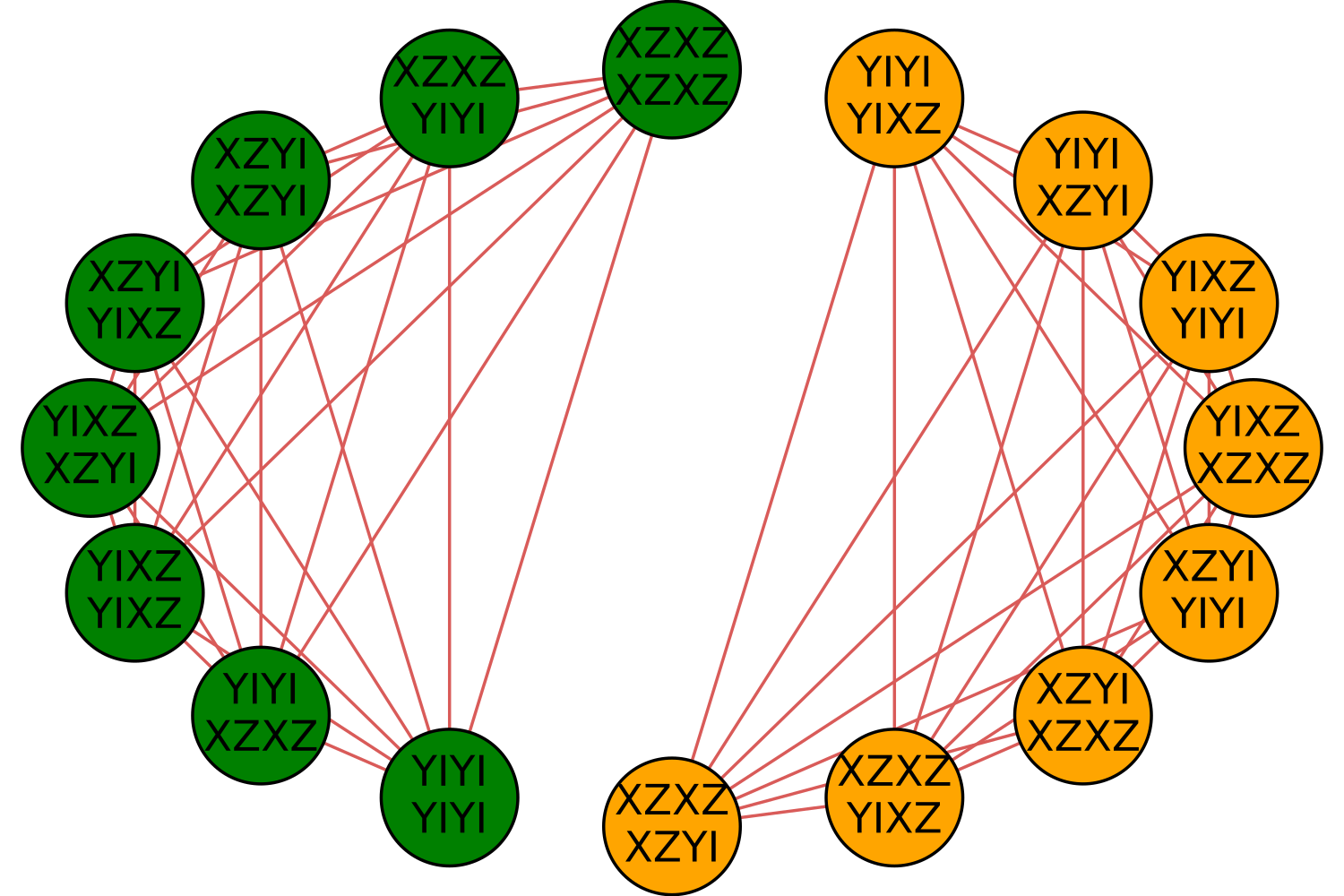}
    \caption{Similar to the Jordan-Wigner case, the 16 relevant Pauli strings in the Parity encoding of $a^{\dagger}_p a^{\dagger}_q a_r a_s$ have a MIN-CLIQUE-COVER of size 2.}
    \label{fig:parity_pqrs_clique}
\end{figure}

For the Parity encoding, we make the transformations:
$$a_p = X_{N-1}...X_{p+1} \frac{X_p Z_{p-1} + i Y_p I_{p-1}}{2}$$
$$a^{\dagger}_p = X_{N-1}...X_{p+1} \frac{X_p Z_{p-1} - i Y_p I_{p-1}}{2}$$

\subsubsection{Double excitation operators}.
WLOG, suppose $p - 1 > q, q - 1 > r, r - 1 > s$. Multiplying out $a_p^{\dagger} a_q^{\dagger} a_r a_s$ we see that the parity encoding creates Pauli strings matching the regular expression:
$$(X_pZ_{p-1}|Y_pI_{p-1}) X_{p-2}...X_{q+1}  (X_qZ_{q-1}|Y_qI_{q-1})...$$
$$...(X_rZ_{r-1}|Y_rI_{r-1}) X_{r-2}...X_{s+1} (X_sZ_{s-1}|Y_sI_{s-1})$$

Only indices $p, p-1, q, q-1, r, r-1, s$, and $s-1$ are relevant for commutativity. Once again expanding the resulting 16 Pauli strings, we see that the commutation graph has a MIN-CLIQUE-COVER of size 2, as depicted in Figure~\ref{fig:parity_pqrs_clique}. Thus, we can again achieve an 8x reduction in the number of partitions by performing simultaneous measurement across these indices. However, note that the simultaneous measurement circuit now involves 8 indices, so it will be more expensive than the simultaneous measurement circuit for the Jordan-Wigner encoding.

\subsubsection{Number and number-excitation operators}
We also again consider the $O(N)$ and $O(N^2)$ operators that are frequent in smaller molecules. The parity encoding on the number and number-excitation operators gives rise to Pauli strings of form $Z_p Z_{p-1}$ and $Z_{p} Z_{p-1} Z_q Z_{q-1}$ respectively. Again, we see that for small molecules, the parity encoding creates a large set of QWC Pauli strings.

\subsection{Bravyi-Kitaev}
The Bravyi-Kitaev coding is asymptotically favorable for Hamiltonian simulation because it requires asymptotically fewer non-$I$ operators per Pauli string by only selecting a subset of indices to perform partial sums needed in the fermion-to-qubit encoding. As a result, every $a_p$ or $a_j^{\dagger}$ term involves a subset of indices ($>p$) that carry the $X$ update, and a subset of the indices ($< j$) that require the phase correction. This complicates the commutation structure of $a_p^{\dagger} a_q^{\dagger} a_r a_s$ and there is not an immediately obvious clique cover strategy--we identify this as an open question.
\section{Circuits for Simultaneous Measurement} \label{sec:circuits_for_simultaneous_measurement}

Once an approximate MIN-COMMUTING-PARTITION solution has been generated, a natural question arises of how to actually perform the necessary simultaneous measurement for each commuting partition. In the case of Naive partitions where each Pauli string is measured separately, the measurement circuit is trivial. In particular, recall from Section~\ref{sec:background} that we simply perform the $H$ and $HS^{\dagger}$ operations on the indices with $X$ or $Y$ respectively, and then we measure every qubit in the Z basis. Thus, we need just $O(N)$ fully-parallelizable single qubit gates; more specifically, we require $k \leq N$ single qubit gates, where $k$ is the number of indices in the Pauli string that equal $X$ or $Y$.

Simultaneous measurement is also similarly straightforward in the case of QWC partitions. Each index of a QWC partition is characterized by a measurement basis. For example, consider the task of simultaneously measuring the two QWC Pauli strings $XIYIZI$ and $IXIYIZ$. We simply apply $H$ to the left two qubits and $HS^{\dagger}$ to the right two qubits. The resulting qubits can all be measured in the standard $Z$ basis, and the corresponding outcomes indicate the $X$, $X$, $Y$, $Y$, $Z$, and $Z$ outcomes as desired. In terms of circuit cost, QWC measurement is essentially identical to Naive measurement: $O(N)$ single qubit gates are required, and the gates are fully parallelizable to constant depth.

While Naive and QWC partition measurements are straightforward, GC partition measurements are nontrivial. We now introduce a circuit synthesis procedure enabling these measurements, and we analyze both the quantum and classical costs of this procedure. To the best of our knowledge, this is the first work explicitly demonstrating how to perform simultaneous measurement in the general case of GC Pauli strings. We implemented our circuit synthesis tool as a Python library and validated it across a wide range of molecular Hamiltonians.

\subsection{Background}

As discussed in Section~\ref{sec:background}, performing a simultaneous measurement amounts to applying a unitary transformation in which the columns of the unitary matrix are the simultaneous eigenvectors of the commuting Pauli strings in the partition. After applying such a transformation and then performing standard Z-basis measurements, the outcomes are mapped directly to measurements of the Pauli strings of interest. One approach to synthesize a simultaneous measurement circuit would be to explicitly compute the matrix of simultaneous eigenvectors and then apply one of many possible unitary decomposition techniques \cite{dawson2005solovay, khaneja2000cartan, ross2014optimal, li2013decomposition, daskin2011decomposition, nakajima2005new} to this matrix. However, this approach is not sufficient for two reasons. First, in general, decomposition techniques trade off between requiring intractable quantum circuit depth, requiring intractable classical compilation time, and yielding only approximations to the desired transformation. Second, and most importantly, these techniques require us to compute the simultaneous eigenvectors and input them to the decomposer. In general, the simultaneous eigenvectors resulting from GC can be fully entangled across all $N$ indices, and they are represented by a $2^N$-sized column vector. The corresponding unitary matrix would be doubly exponentially sized in $N$, erasing any potential quantum advantage.

With this in mind, it is clear that any decomposition technique must avoid explicitly computing eigenvectors and writing out exponentially sized unitary matrices. Fortunately, the stabilizer formalism---typically applied to quantum error correction---provides us such a mechanism. Before proceeding, we note that our work is built upon the language of stabilizers introduced in \cite{gottesman1997stabilizer} and expanded upon in \cite{aaronson2004improved}. While these two papers were applied to error correction and quantum simulation, the core techniques also apply to our use case. Also, \cite{seyfarth2011construction} and \cite{seyfarth2019cyclic} leverage these stabilizer techniques to perform MUB measurements. Our circuit constructions are drawn from these two papers as well as \cite{zheng2018depth}, but stem from a different context and end goal.

\subsection{An Example: $\{XX, YY, ZZ\}$}
\begin{figure}[h!]
$$
\Qcircuit @C=1em @R=1.6em {
\lstick{} & \qw & \ctrl{1} & \gate{H}  & \measureZ \\
\lstick{} & \qw & \targ    & \qw       & \measureZ
\inputgroupv{1}{2}{.8em}{1.6em}{\ket{\psi}}\\
}$$
    \caption{Bell basis measurement circuit that simultaneously measures $XX$, $YY$, and $ZZ$ on the $\ket{\psi}$ state. After application of these two gates, the measurements of the top and bottom qubits correspond to outcomes for $XX$ and $ZZ$ respectively. The $YY$ outcome is obtained from $YY = - (XX) (ZZ)$.}
    \label{fig:bell_basis_measurement_circuit}
\end{figure}
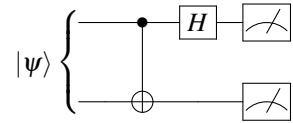

We begin with a well-known example. Consider the task of trying to simultaneously measure $XX, YY,$ and $ZZ$, a GC (but not QWC) partition. The simultaneous eigenvectors of these Pauli strings are known as the four \textit{Bell states}:
$$
\ket{\Phi^+} = \frac{\ket{00} + \ket{11}}{\sqrt{2}}, \qquad
\ket{\Phi^-} = \frac{\ket{00} - \ket{11}}{\sqrt{2}},
$$
$$
\ket{\Psi^+} = \frac{\ket{01} + \ket{10}}{\sqrt{2}}, \qquad
\ket{\Psi^-} = \frac{\ket{01} - \ket{10}}{\sqrt{2}}
$$
These eigenvectors are linearly independent and span all possible 2-qubit states---hence, they are a basis. Unlike the vectors in the standard computational basis of $\{\ket{00}, \ket{01}, \ket{10}, \ket{11}\}$, the eigenvectors in the Bell basis feature entanglement between the two qubits. As a result, measurement in the Bell basis requires interaction between the two qubits, unlike the the Naive and QWC measurements described previously. The quantum circuit in Figure~\ref{fig:bell_basis_measurement_circuit} is a well-known circuit that performs Bell basis measurement, i.e. simultaneous measurement of $XX$, $YY$, and $ZZ$.

To understand why this circuit measures $XX$ and $ZZ$ (and also $YY = - (XX) (ZZ)$), we observe that our ultimate goal is to transform a target measurement of $[XX, ZZ]$ into $[ZI, IZ]$---the latter captures the outcomes we actually measure directly via standard Z-basis measurement. An important background result is that after applying some unitary operation $U$, a target measurement of $M$ on the original state has become equivalent to a measurement of $UMU^{\dagger}$ \cite{gottesman1997stabilizer, nielsen2010quantum} on the new state. This is known as \textit{unitary conjugation}.

In the Bell basis measurement circuit, we first apply $U = CNOT$. By computing $UMU^{\dagger}$ we can see that target measurements of $[XX, ZZ]$ are transformed under conjugation to measurements of
$$[XX, ZZ] \xrightarrow[\text{$U = CNOT$}]{\text{$UMU^{\dagger}$}} [U XX U^{\dagger}, U ZZ U^{\dagger}] = [XI, IZ].$$
Finally, after applying the Hadamard gate on the top qubit, the measurements are transformed to
$$[XI, IZ] \xrightarrow[\text{$U = H \otimes I$}]{\text{$UMU^{\dagger}$}} [U XI U^{\dagger}, U IZ U^{\dagger}] = [ZI, IZ].$$
Thus, this $CNOT$, $H \otimes I$ gate sequence performs the desired transformation of rotating a measurement of $[XX, ZZ]$ into the computational basis, $[ZI, IZ]$. The ordering of the elements is important and indicates that measurement of the top qubit ($ZI$) corresponds to the $XX$ outcome and measurement of the bottom qubit ($IZ$) corresponds to the $ZZ$ outcome. As mentioned previously, $YY$ follows as $- (XX) (ZZ)$.

\subsection{Stabilizer Matrices}

In order to consider the general case, we now switch to the formalism of stabilizer matrices. Our notation and terminology is similar to previous work \cite{gottesman1997stabilizer, aaronson2004improved, seyfarth2011construction, seyfarth2019cyclic, zheng2018depth}, with some deviations for clarity. Within the stabilizer formalism, every $N$-qubit Pauli string maps to a $2N$-entry column vector. The top $N$ entries indicate whether each corresponding index `contains' a $Z$. The bottom $N$ entries correspond to $X$'s. The $Y$ Pauli matrix corresponds to having a 1 in both the $Z$ and $X$ entries, since $Y = iZX$. The stabilizer matrix for a list of Pauli strings is simply the concatenation of the column vectors. As an instructive example, the stabilizer matrix for $[XXX, YYY, ZZZ, XYZ]$ is:
$$
\begin{pmatrix}
0 & 1 & 1 & 0 \\
0 & 1 & 1 & 1 \\
0 & 1 & 1 & 1 \\
1 & 1 & 0 & 1 \\
1 & 1 & 0 & 1 \\
1 & 1 & 0 & 0
\end{pmatrix}
$$

For convenience and clarity, we will refer to the top $N$ rows as the \textit{$Z$-matrix} and the bottom $N$ rows as the \textit{$X$-matrix}. Recall that our goal is to transform a target set of Pauli strings for simultaneous measurement into the computational basis measurements, $[ZII...I, IZI...I, ..., III...Z]$. We see that the stabilizer matrix for this computational basis simply has an $N \times N$ Identity as the $Z$-matrix and all zeroes in the $X$-matrix.

We now seek a procedure to transform the target stabilizer matrix into this computational basis stabilizer matrix. To see how to accomplish such a transformation, it is useful to know unitary conjugation relationships for a basic gate set. Table~\ref{tab:single_qubit_conjugation} and Table~\ref{tab:two_qubit_conjugation} list the unitary conjugations of important Pauli strings for 1- and 2- qubit unitary gates respectively.

\begin{table}[H]
\centering
\begin{tabular}{l||c|c}
      & $UZU^{\dagger}$ & $UXU^{\dagger}$ \\ \hline \hline
$U=H$ & $X$             & $Z$             \\ \hline
$U=S$ & $Z$             & $Y$            
\end{tabular}
\caption{Result of conjugation of $Z$ and $X$ by single qubit gates $U = $ $H$ or $S$. Note that $H$ can be thought of as a ``NOT gate'' between $X$ and $Z$. The $S$ (phase) gate does not affect $Z$, but does transform $X$ into $Y$.}
\label{tab:single_qubit_conjugation}
\end{table}

\begin{table}[H]
\centering
\begin{tabular}{l||c|c|c|c}
         & $UZIU^{\dagger}$ & $UIZU^{\dagger}$ & $UXIU^{\dagger}$ & $UIXU^{\dagger}$ \\ \hline \hline
$U=CNOT$ & $ZI$             & $ZZ$              & $XX$             & $IX$             \\ \hline
$U=CZ$   & $ZI$             & $IZ$              & $XZ$             & $ZX$             \\ \hline
$U=SWAP$ & $ZI$             & $ZI$             & $IX$             & $XI$
\end{tabular}
\caption{Result of conjugation of $ZI$, $IZ$, $XI$, or $IX$ by two qubit gates $U = $ $CNOT$, $CZ$, or $SWAP$.}
\label{tab:two_qubit_conjugation}
\end{table}

Based on these tables, we can interpret the action of each of these unitaries on a stabilizer matrix. These rules can be verified directly from the tables and are also explained in \cite{aaronson2004improved, seyfarth2019cyclic}.
\begin{itemize}
    \item $H$ on the $i$th qubit swaps the $i$th and $i+N$th row of the stabilizer matrix (i.e. swaps between corresponding rows of the $Z$- and $X$- matrices). It is helpful to think of $H$ as a "NOT gate" that flips $Z$ and $X$ measurements.
    \item $S$ on the $i$th qubit sets the $(i, i)$ diagonal entry in the $Z$-matrix to 0.
    \item $CNOT$ controlled on $i$th qubit and targeted on the $j$th qubit adds the $j$th row to the $i$th row and adds $i + N$th row to the $j + N$th row. All additions are performed modulo 2.
    \item $CZ$ between the $i$ and $j$th qubits sets the $(i, j)$ and $(j, i)$ symmetric off-diagonal entries of the $Z$-matrix to 0.
    \item $SWAP$ between the $i$ and $j$th qubits swaps the $i$ and $j$th rows of both the $Z$ and $X$ matrices. This can be seen from the fact that $SWAP = (CNOT)(NOTC)(CNOT)$ and two rows can be swapped with three alternating binary additions.
\end{itemize}

\subsection{Circuit Synthesis Procedure}
We now have the tools we need for circuit synthesis, which amounts to transforming the stabilizer matrix for a commuting family of Pauli strings into the computational basis stabilizer matrix (which has Identity for the $Z$-matrix and zeros for the $X$-matrix). For simplicity, we describe the procedure for the case when the partition of $N$-qubit Pauli strings is complete and contains $N$ linearly independent elements. This is the hardest case---if the partition is incomplete, the measurement procedure is similar but has more slack, because at least 1 of the qubits will not need to be measured.

\begin{algorithm}
\SetAlgoLined
\SetKwInOut{Input}{input}\SetKwInOut{Output}{output}
\Input{ $\{P_i\}$, a complete GC family of Pauli strings}
\Output{ Circuit for simultaneous measurement of $\{P_i\}$}
\BlankLine
$M \in F_2^{2N \times N} \leftarrow$ basis of $\{P_i\}$\;
Full-rankify $Z$-matrix by applying $H$ gates\;
Gaussian eliminate $X$-matrix using CNOT \& SWAP gates\;
\For{each diagonal element in $Z$-matrix}{
\lIf{element is 1}{apply $S$ to corresponding qubit}
}
\For{each element below diagonal of $Z$-matrix}{
\lIf{element is 1}{apply $CZ$ to the row-col qubits}
}
Apply $H$ to each qubit\;
Measure each qubit\;
\caption{Circuit synthesis for sim. measurement}
\label{alg:circuit_synthesis}
\end{algorithm}

The circuit synthesis procedure is described in Algorithm~\ref{alg:circuit_synthesis}. To develop its intuition, we demonstrate its application to the problem of simultaneously measuring $[IYX, ZZZ, XIX, ZXY]$, which is a GC (but not QWC) family. We initialize the algorithm by setting the stabilizer matrix to a basis of this partition. Note that the fourth term is linearly dependent on the first three, so we exclude it to yield such a basis; in general, we use Gaussian elimination to perform this distillation of the Pauli strings into a basis. The stabilizer matrix for this resulting list of Pauli strings, $[IYX, ZZZ, XIX]$, is:
\vspace*{-\baselineskip}
$$
\begin{pmatrix}
0 & 1 & 0 \\
1 & 1 & 0 \\
0 & 1 & 0 \\
0 & 0 & 1 \\
1 & 0 & 0 \\
1 & 0 & 1
\end{pmatrix}$$

The first step of the simultaneous measurement circuit synthesis is to apply $H$ gates as needed to transform the $X$-matrix to have full rank (it is currently only rank 2). Such a transformation is always possible and can be found efficiently by Gaussian elimination \cite[Lemma 6]{aaronson2004improved}. In this case, applying $H$ to the first qubit swaps the first and fourth rows of the stabilizer matrix, yielding an $X$-matrix of full rank 3:

\begin{table}[H]
\vspace*{-\baselineskip}
\begin{tabular}{ m{1cm} m{2cm} m{1cm} m{3cm} }
& $$\Qcircuit @C=1em @R=1.0em {
& \gate{H} & \qw \\
& \qw & \qw \\
& \qw & \qw
}$$ & $\rightarrow$ & $\begin{pmatrix} 0 & 0 & 1 \\ 1 & 1 & 0 \\ 0 & 1 & 0 \\ 0 & 1 & 0 \\ 1 & 0 & 0 \\ 1 & 0 & 1 \end{pmatrix}$ \\
\end{tabular}
\end{table}

Now that the $X$-matrix is of full rank, we can apply standard Gaussian elimination to row reduce it into the Identity matrix. The CNOT and SWAP gates give us the elementary row operations needed: add one row to another and swap rows. In this example, the $X$-matrix can be row reduced to the identity by first adding its second row to the third row, and then swapping the first and second rows. Breaking this down, we first observe the effect of the CNOT on the stabilizer matrix: 
\begin{table}[H]
\vspace*{-\baselineskip}
\begin{tabular}{ m{1cm} m{2cm} m{1cm} m{3cm} }
& $$\Qcircuit @C=1em @R=1.0em {
& \qw & \qw \\
& \ctrl{1} & \qw \\
& \targ & \qw
}$$ & $\rightarrow$ & $\begin{pmatrix}
0 & 0 & 1 \\ 1 & 0 & 0 \\ 0 & 1 & 0 \\
0 & 1 & 0 \\ 1 & 0 & 0 \\ 0 & 0 & 1
\end{pmatrix}$ \\
\end{tabular}
\end{table}

And finally the SWAP completes the row reduction, leaving the $X$-matrix as the identity:
\begin{table}[H]
\vspace*{-\baselineskip}
\begin{tabular}{ m{1cm} m{2cm} m{1cm} m{3cm} }
& $$\Qcircuit @C=1em @R=1.0em {
& \qswap & \qw \\
& \qswap \qwx & \qw \\
& \qw & \qw \\
}$$ & $\rightarrow$ & $\begin{pmatrix}
1 & 0 & 0 \\ 0 & 0 & 1 \\ 0 & 1 & 0 \\
1 & 0 & 0 \\ 0 & 1 & 0 \\ 0 & 0 & 1
\end{pmatrix}$ \\
\end{tabular}
\end{table}

Notice that the CNOT and SWAP also affected the $Z$-matrix, which is now a symmetric matrix; this is guaranteed to occur \cite{seyfarth2019cyclic}. Now our desired transformation is almost complete. The on-diagonal 1 is erased with $S$ on the first qubit, and the two off-diagonal 1s are erased with a $CZ$ between the second and third qubits. These two operations have no effect on the $X$-matrix:
\begin{table}[H]
\vspace*{-\baselineskip}
\begin{tabular}{ m{1cm} m{2cm} m{1cm} m{3cm} }
& $$\Qcircuit @C=1em @R=1.0em {
& \gate{S} & \qw \\
& \ctrl{1} & \qw \\
& \control \qw & \qw
}$$ & $\rightarrow$ & $\begin{pmatrix}
0 & 0 & 0 \\ 0 & 0 & 0 \\ 0 & 0 & 0 \\
1 & 0 & 0 \\ 0 & 1 & 0 \\ 0 & 0 & 1
\end{pmatrix}$ \\
\end{tabular}
\end{table}

Finally, we apply an $H$ to each qubit, which swaps the $Z$- and $X$- matrices, leaving us in the computational basis stabilizer matrix, as desired:
\begin{table}[H]
\vspace*{-\baselineskip}
\begin{tabular}{ m{1cm} m{2cm} m{1cm} m{3cm} }
& $$\Qcircuit @C=1em @R=1.0em {
& \gate{H} & \qw \\
& \gate{H} & \qw \\
& \gate{H} & \qw
}$$ & $\rightarrow$ & $\begin{pmatrix}
1 & 0 & 0 \\ 0 & 1 & 0 \\ 0 & 0 & 1 \\
0 & 0 & 0 \\ 0 & 0 & 0 \\ 0 & 0 & 0
\end{pmatrix}$ \\
\end{tabular}
\end{table}

The full circuit and resulting transformation is shown below:
\begin{table}[H]
\vspace*{-\baselineskip}
\begin{tabular}{m{1.5cm} m{0.2cm} m{2.8cm} m{0.2cm} m{1cm} }
$\begin{pmatrix}
0 & 1 & 0 \\ 1 & 1 & 0 \\ 0 & 1 & 0 \\
0 & 0 & 1 \\ 1 & 0 & 0 \\ 1 & 0 & 1\end{pmatrix}$ & $\rightarrow$ & $$\Qcircuit @C=1em @R=1.0em {
& \gate{H} & \qswap         & \gate{S}     & \gate{H} & \qw \\
& \ctrl{1} & \qswap \qwx    & \ctrl{1}     & \gate{H} & \qw \\
& \targ    & \qw             & \control \qw & \gate{H} & \qw
}$$ & $\rightarrow$ & $\begin{pmatrix}
1 & 0 & 0 \\ 0 & 1 & 0 \\ 0 & 0 & 1 \\
0 & 0 & 0 \\ 0 & 0 & 0 \\ 0 & 0 & 0
\end{pmatrix}$ \\
\end{tabular}
\end{table}

\subsection{Circuit Complexity}
The efficiency of Algorithm~\ref{alg:circuit_synthesis} and the overarching stabilizer formalism stems from the fact that the stabilizer matrices are of size $2N \times N$, and all manipulations are on this tractably-sized matrix. This averts the exponential cost that manipulating simultaneous eigenvectors would entail. In terms of classical cost, the synthesis tool is fast because its slowest step is the Gaussian elimination, which has time complexity of $O(N^3$) \cite{farebrother1988linear}.

The actual circuit produced by the synthesis procedure requires only $O(N^2)$ gates in the worst case, as also noted in related results \cite{seyfarth2019cyclic, zheng2018depth}. This follows because the Gaussian elimination can require $O(N^2)$ elementary row operations, which entails $O(N^2)$ CNOT gates. The erasure of off-diagonal elements in the $Z$-matrix also requires $O(N^2)$ CZ gates.

While the $O(N^2)$ gate count for GC measurement is worse scaling than the $O(N)$ gate count for Naive or QWC measurement, we emphasize that the measurement circuit is preceded by an ansatz preparation circuit that dominates gate counts and depth. In particular, the UCCSD ansatz has $O(N^4)$ gate count and $O(N^3)$ depth after parallelization. Therefore, the cost of simultaneous measurement is asymptotically insignificant. As discussed, we base our studies on UCCSD because the Coupled Cluster approach is the gold standard for quantum computational chemistry \cite{mcardle2018quantum, bartlett2007coupled}. Moreover, UCCSD has shown experimental and theoretical promise, unlike hardware-driven ansatz, which were shown to suffer from ``barren plateaus'' in the optimization landscape \cite{mcclean2018barren, mcardle2018quantum}. Even in the case of other non-hardware-driven ansatzes, gate counts and depths generally scale at least as $N^3$ in order to achieve high accuracy. Thus, the quadratic cost of GC measurement appears to be benign.

We also underscore that the $O(N^2)$ gate count scaling of simultaneous measurement is a worst case scenario, where our partition is dominated by GC-but-not-QWC edges. In practice, this is not the case and we see QWC on many, if not most indices. For example, in the linear-time MIN-COMMUTING-PARTITION 8x approximations presented in Section~\ref{sec:linear_time_partitioning} only a constant (4 or 8) number of Pauli string indices have a GC-but-not-QWC relationship in the simultaneous measurements. The remaining $N-4$ or $N-8$ Pauli string indices are QWC. Thus, under this MIN-COMMUTING-PARTITION approximation, the simultaneous measurement circuit gate count is still $O(N)$ and the depth is still parallelizable to $O(1)$.

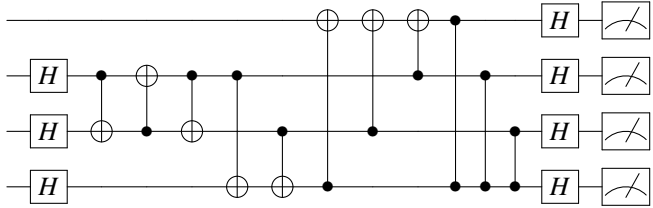
\begin{figure}[h!]
    \centering
\Qcircuit @C=.89em @R=.7em {
& \qw      & \qw      & \qw       & \qw      & \qw      & \qw      & \targ     & \targ     & \targ     & \ctrl{3}     & \qw          & \qw          & \gate{H} & \measureZ \\
& \gate{H} & \ctrl{1} & \targ     & \ctrl{1} & \ctrl{2} & \qw      & \qw       & \qw       & \ctrl{-1} & \qw          & \ctrl{2}     & \qw          & \gate{H} & \measureZ \\
& \gate{H} & \targ    & \ctrl{-1} & \targ    & \qw      & \ctrl{1} & \qw       & \ctrl{-2} & \qw       & \qw          & \qw          & \ctrl{1}     & \gate{H} & \measureZ \\
& \gate{H} & \qw      & \qw       & \qw      & \targ    & \targ    & \ctrl{-3} & \qw       & \qw       & \control \qw & \control \qw & \control \qw & \gate{H} & \measureZ
}
\caption{Simultaneous measurement circuit generated by our software for the green 8-clique in Figure~\ref{fig:jw_pqrs_clique_circuit}. It transforms the measurements of $XXXX, XXYY, XYXY, YXXY$ (which is a basis for the Pauli strings in the green 8-clique) to measurements of $ZIII, IZII, IIZI, IIIZ$.}
    \label{fig:jw_pqrs_clique_circuit}
\end{figure}

For reference, we show in Figure~\ref{fig:jw_pqrs_clique_circuit} the simultaneous measurement circuit for the 4 GC-but-not-QWC qubits in the Pauli partition for the Jordan-Wigner transformation. Specifically, this measurement circuit is used to measure the green 8-clique in Figure~\ref{fig:jw_pqrs_clique}. The other $N-4$ qubits are QWC and require single-qubit gates for measurement---this is why the simultaneous measurement gate complexity is still just $O(N)$.

\subsection{Measurement Circuit Optimizations}
While the circuit synthesis procedure in Algorithm~\ref{alg:circuit_synthesis} yields a correct simultaneous measurement circuit, it is not necessarily the most optimal circuit possible. For instance, in Figure~\ref{fig:jw_pqrs_clique_circuit}, the $SWAP$ (implemented as 3 $CNOT$s) between qubits 2 and 3 can be omitted from the circuit and instead implemented by swapping their subsequent gates, and then accounting for the SWAP classically after the measurements are performed. In other words, the SWAPs in our circuit constructions can be accomplished by simple classical re-labeling of qubit indices.

We also observe that many gates can be parallelized. For example, the depth of Figure~\ref{fig:jw_pqrs_clique_circuit} can be reduced by parallelizing the execution of the $CZ$ gates with the execution of the $CNOT$ gates.
\section{Benchmark Results} \label{sec:benchmark_results}
We tested the performance of our simultaneous measurement strategies in Section~\ref{sec:min_clique_cover} on multiple molecular benchmarks, whose Hamiltonians we obtained via OpenFermion \cite{mcclean2017openfermion}. Our benchmark results encompass both the reduction in number of partitions relative to Naive, as well as the classical computation runtime required to produce the partitioning.

\begin{figure}[h]
    \centering
    \includegraphics[width=0.45\textwidth]{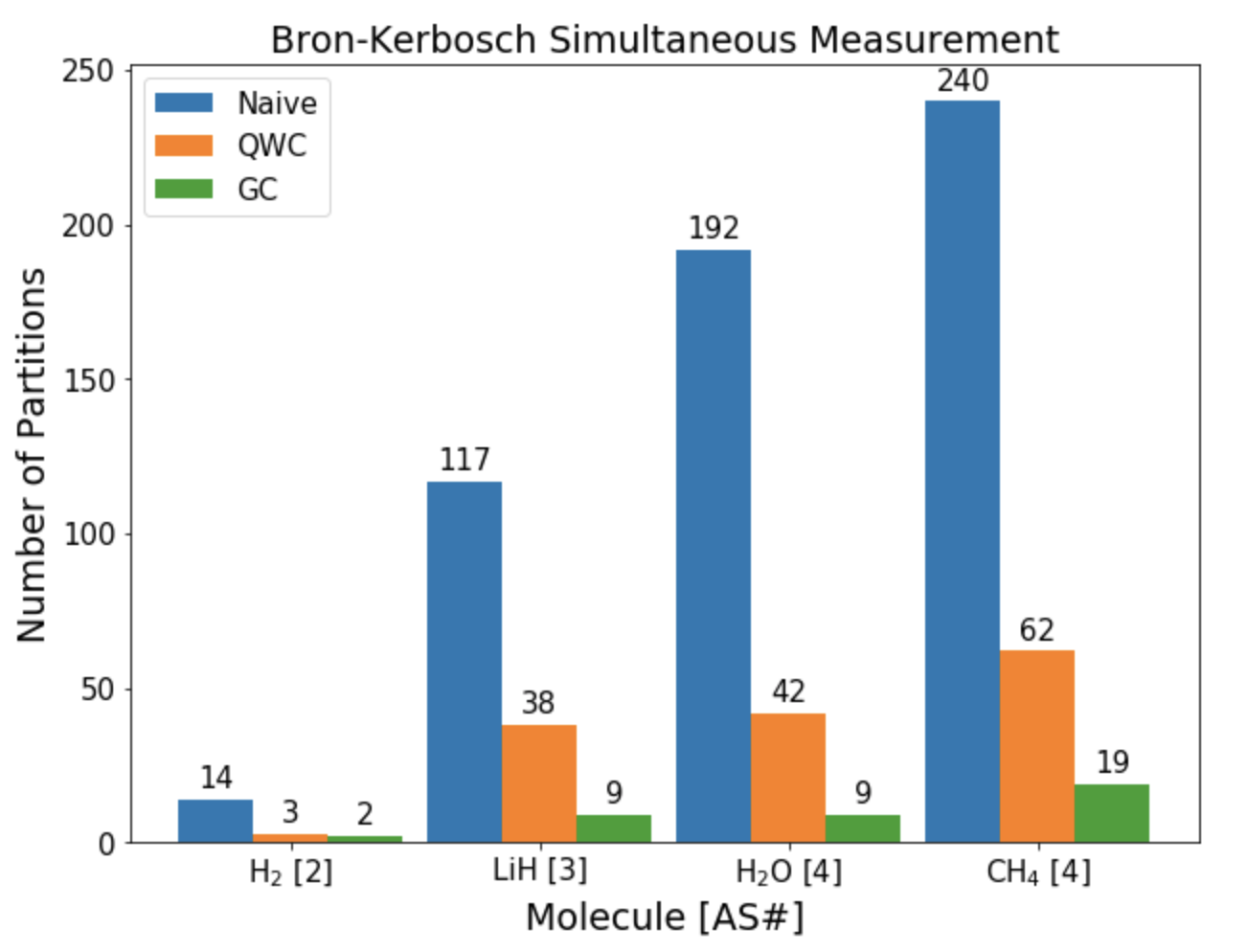}
    \caption{Number of QWC and GC partitions (which we are attempting to \textit{minimize}) generated by Bron-Kerbosch for four representative molecules. AS\# indicates the number of active spaces for the molecular Hamiltonian.}
    \label{fig:bk_molecules}
\end{figure}

As mentioned, in Section~\ref{sec:min_clique_cover}, the Bron-Kerbosch based MIN-CLIQUE-COVER approximation has exponential worst case runtime and should thus be considered a soft bound on the optimality of partitions produced by other graph approximation algorithms. Figure~\ref{fig:bk_molecules} indicates the performance of Bron-Kerbosch in terms of number of commuting partitions (cliques) found using both QWC and GC edges, in comparison to the Naive VQE implementation in which each Pauli string is in a singleton partition. The improvement from Naive to QWC is consistently about 4-5x---a significant reduction especially considering that QWC measurement is cheap. The improvement from Naive to GC ranges from 7x to 12x from H\textsubscript{2} to CH\textsubscript{4} (methane). This suggests that the state preparation cost reduction factor from GC partitioning improves for larger molecules.

\begin{figure}[h]
    \centering
    \includegraphics[width=0.45\textwidth]{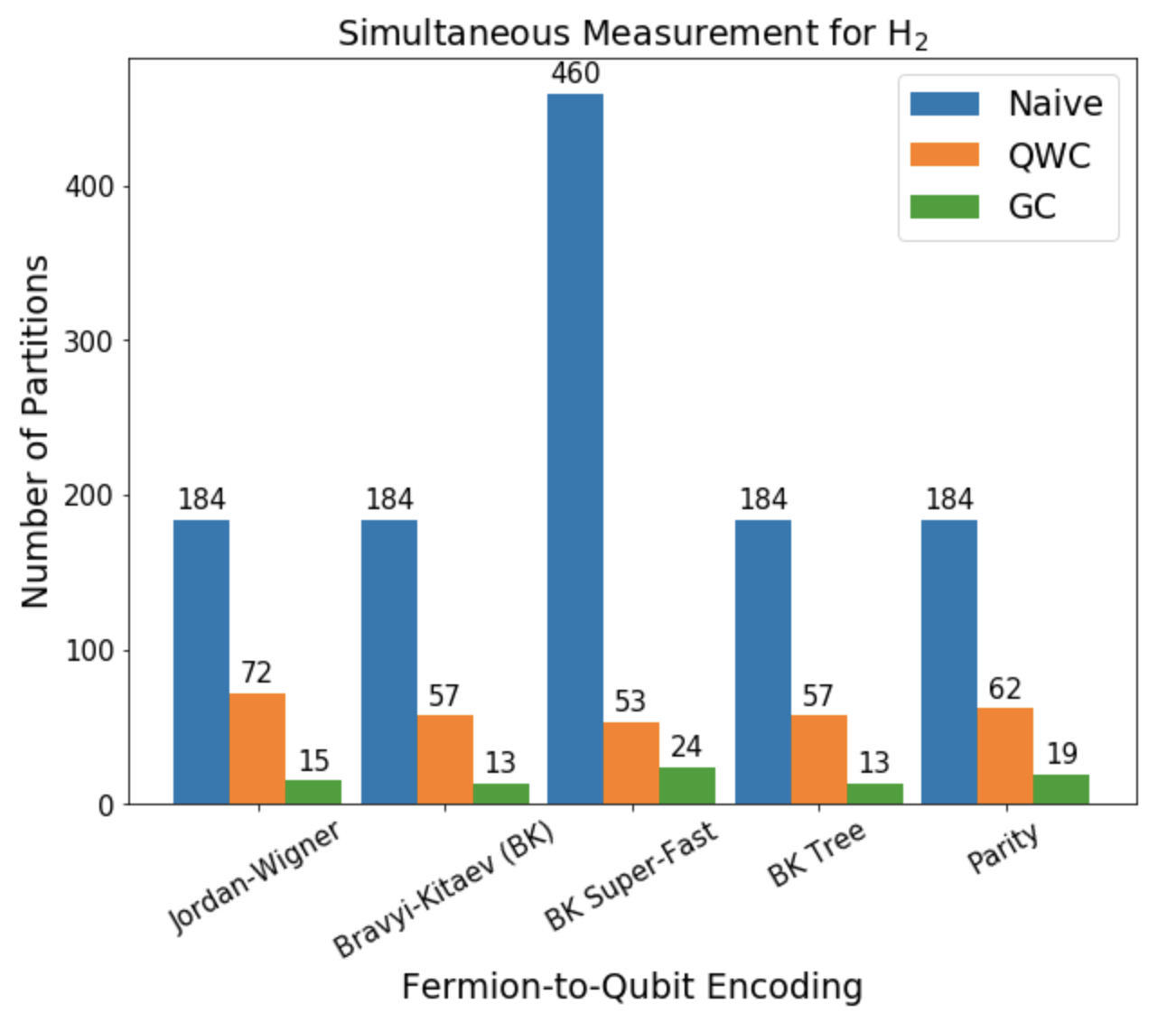}
    \caption{Number of QWC and GC partitions generated by Bron-Kerbosch for the H\textsubscript{2} molecule, under different fermion-to-qubit encodings.}
    \label{fig:bk_encodings}
\end{figure}

\begin{figure}[h]
    \centering
    \includegraphics[width=0.45\textwidth]{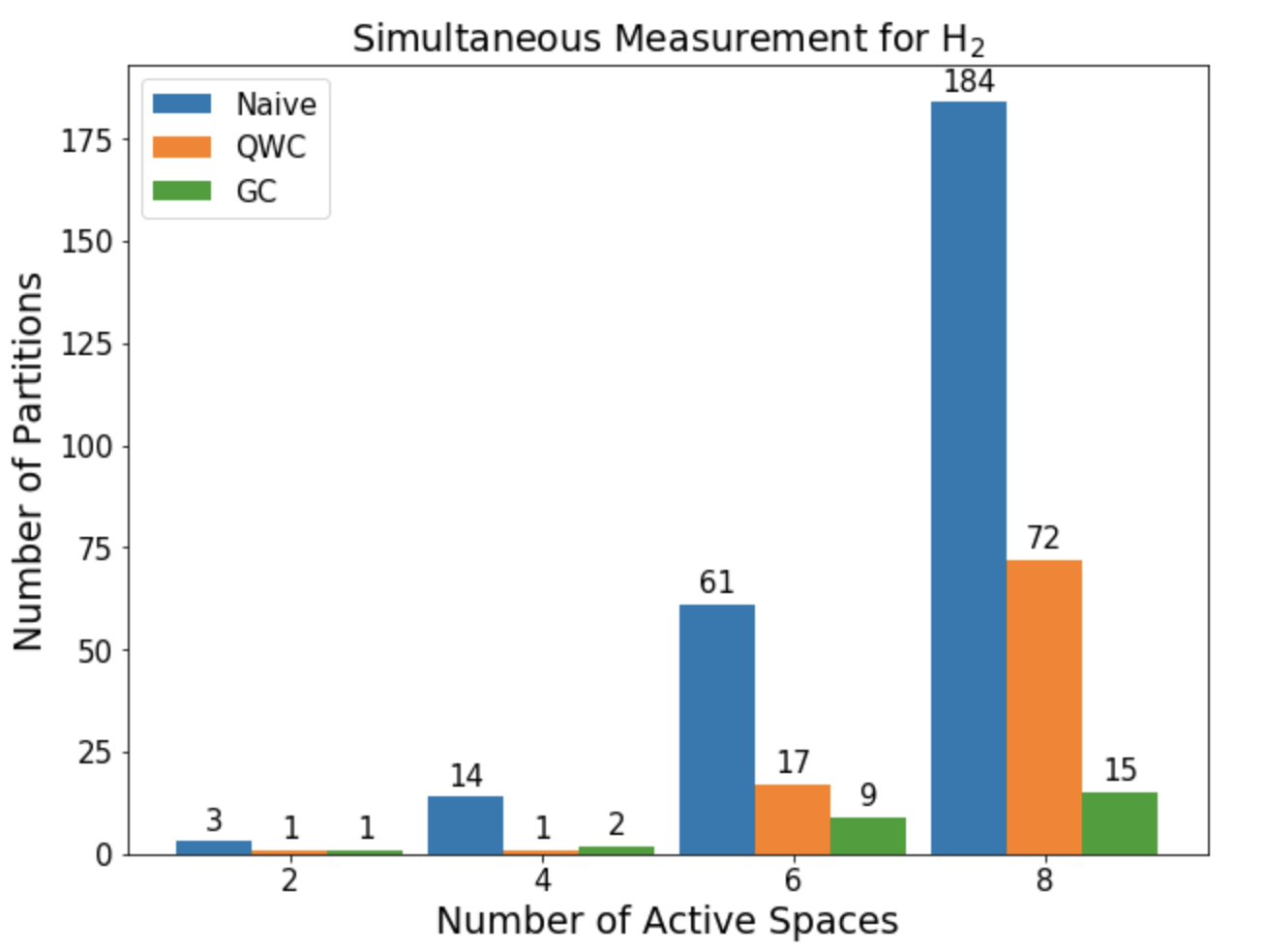}
    \caption{Number of QWC and GC partitions generated by Bron-Kerbosch for the H\textsubscript{2} molecule, under different numbers of active spaces.}
    \label{fig:bk_activespaces}
\end{figure}

Figure~\ref{fig:bk_encodings} and Figure~\ref{fig:bk_activespaces} examine partitioning efficacy when we vary the qubit encodings and the number of active spaces considered for the H\textsubscript{2} molecule. Across the qubit encodings, performance is roughly consistent with a 3x improvement from QWC partitions and a 10x improvement from GC partitions. We do note one outlier in that the performance is particularly promising for the Brayvi-Kitaev Super-Fast encoding \cite{bravyi2002fermionic}, which achieves a 20x reduction in the number of partitions from Naive to GC. Across the varying active spaces, we again see evidence that the GC partitioning advantage scales with Hamiltonian size, ranging from 3x to 12x as the number of active spaces is increased. This is important and encouraging, because prior work demonstrated that a relatively large number of active spaces are needed to achieve chemical accuracy \cite{barkoutsos2018quantum}.

\begin{figure}[h]
    \centering
    \includegraphics[width=0.45\textwidth]{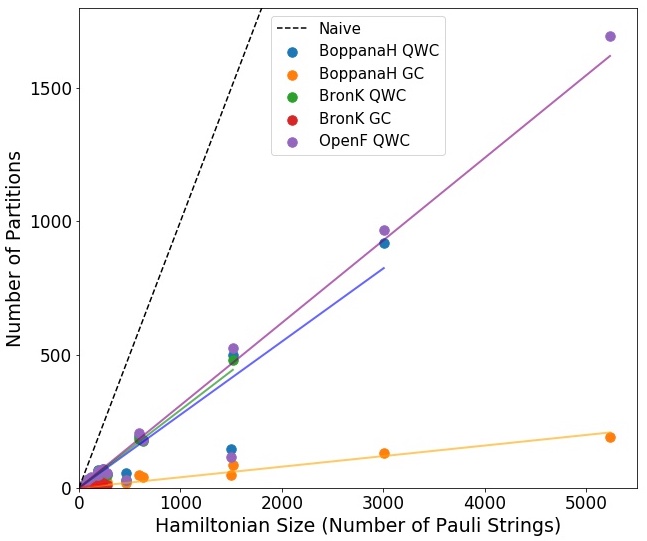}
    \caption{Number of partitions found for each algorithm-benchmark pair. Under Naive measurement, the number of partitions would exactly equal the Hamiltonian size (number of Pauli strings). Thus, these techniques all achieve a 4-20x reduction in state preparations and measurements relative to the Naive strategy.}
    \label{fig:clique_scaling_large}
\end{figure}

\begin{figure}[h]
    \centering
    \includegraphics[width=0.45\textwidth]{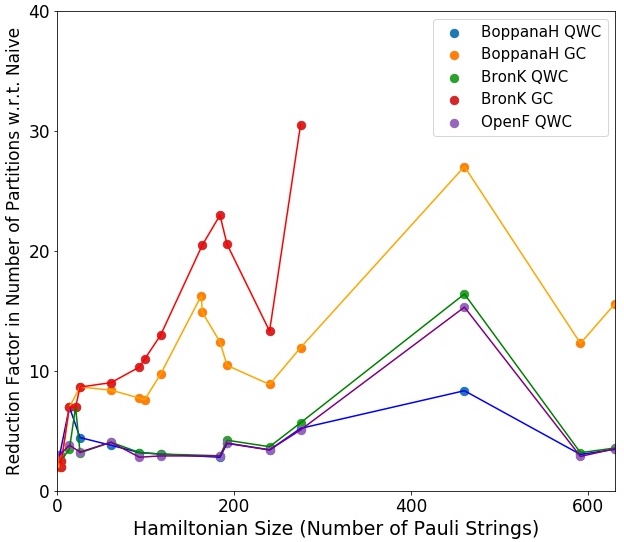}
    \caption{Factor of improvement (which we are attempting to \textit{maximize}) over Naive for each of the algorithms benchmarked for Hamiltonian sizes up to 630 terms.}
    \label{fig:clique_scaling_small}
\end{figure}

Along with the Bron-Kerbosch approximations as a loose upper bound on the expected partitioning optimality, we also benchmarked another MIN-CLIQUE-COVER approximation: the Boppana-Halld\'{o}rsson algorithm, applied to both QWC- and GC- edge graphs. In addition, we also benchmarked with the QWC partitioning heuristic provided by the OpenFermion  electronic structure package. We tested each of these algorithms on problem sizes ranging from 4 to 5237 terms in the molecular Hamiltonian. These Hamiltonians correspond to the H\textsubscript{2}, LiH, H\textsubscript{2}O, and CH\textsubscript{4} molecules with varying numbers of active spaces. We recorded both the number of partitions generated and the runtime for each algorithm-benchmark pair. Figure~\ref{fig:clique_scaling_large} shows the number of partitions generated for Hamiltonians with up to 5237 Pauli strings. Note that some of the benchmarks were unable to be run due to prohibitive runtime costs on the order of days (e.g. Bron-Kerbosch for $|H| > 1519$ Pauli strings). Figure~\ref{fig:clique_scaling_small} shows a zoom-in for molecules with up to 630 Pauli strings; the y-axis now shows the reduction factor in number of partitions. The plots generally align with our expectations: GC leads to much more optimal partitioning than QWC (recall the arguments in Section~\ref{subsec:qwc_vs_gc}, and Bron-Kerbosch GC achieves the fewest number of partitions generated although Boppana-Halld\'{o}rsson GC has comparable optimality. Among the QWC methods, we consistently see 3-4x reductions in number of partitions over Naive separate measurements, and our Boppana-Halld\'{o}rsson QWC algorithm marginally outperforms the OpenFermion heuristic.

\begin{figure}[h]
    \centering
    \includegraphics[width=0.45\textwidth]{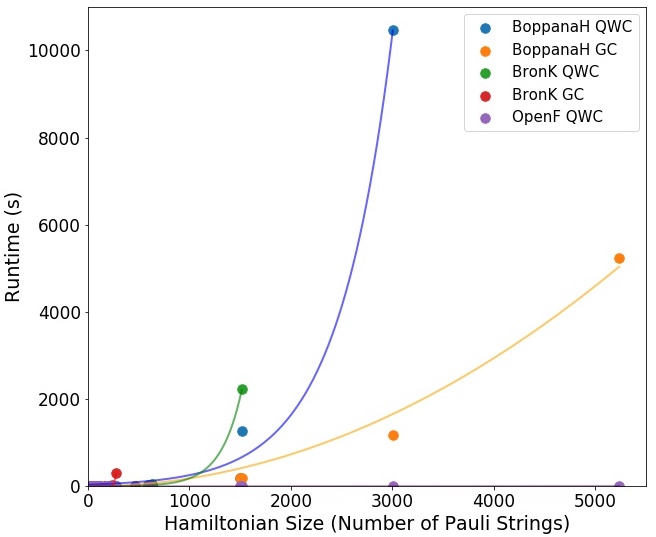}
    \caption{Classical computer runtimes for each partitioning algorithm + benchmark pair. Bron-Kerbosch has exponential and Boppana-Halld\'{o}sson has quadratic runtime scaling. This partitioning step runs as a compilation procedure before the actual quantum invocations of VQE.}
    \label{fig:time_scaling_large}
\end{figure}

\begin{figure}[h]
    \centering
    \includegraphics[width=0.45\textwidth]{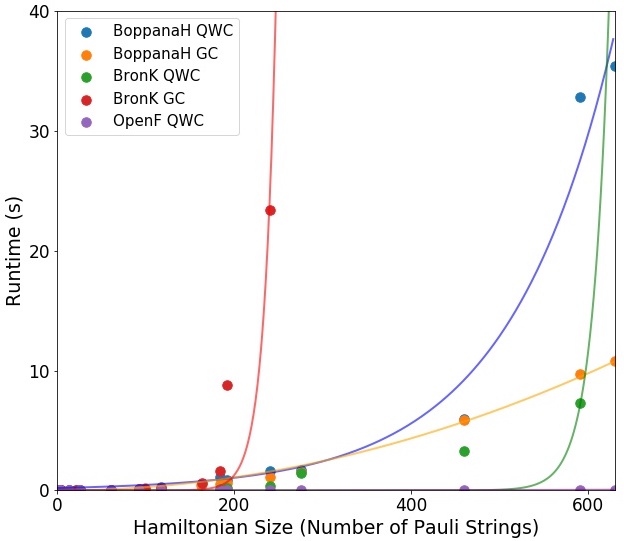}
    \caption{Zoom-in of Figure~\ref{fig:time_scaling_large} for Hamiltonian sizes up to 630 terms.}
    \label{fig:time_scaling_small}
\end{figure}

Figure~\ref{fig:time_scaling_large} plots the wall clock runtimes for each of the algorithm-benchmark pairs; Figure~\ref{fig:time_scaling_small} focuses on the 0 -- 630 Hamiltonian size range. These plots corroborate the exponential worst-case scaling of Bron-Kerbosch and suggest quadratic runtime scaling for the Boppana-Halld\'{o}sson algorithm. OpenFermion's function is clearly the fastest of the algorithms explored, but is also consistently the worst approximation to the MIN-COMMUTING-PARTITION. 

\section{Experimental Results} \label{sec:experimental_results}
We validated our techniques with a proof of concept demonstration by experimentally replicating a recent result \cite{dumitrescu2018cloud}: ground state energy estimation of deuteron, the nucleus of an uncommon isotope of hydrogen. We performed our experiments via the IBM Q Tokyo 20-qubit quantum computer \cite{ibmqtokyo}, which is cloud accessible.

Following \cite{dumitrescu2018cloud}, deuteron can be modeled with a 2-qubit Hamiltonian spanning 4 Pauli strings\footnote{There is also an $II$ term, but this doesn't actually require any measurement---it just adds a constant offset to the Hamiltonian.}: $IZ$, $ZI$, $XX$, and $YY$. Under Naive measurement, each Pauli string is measured in a separate partition. Under GC, we can partition into just two commuting families: $\{ZI, IZ\}$ and $\{XX, YY\}$. Recall that the former partition is QWC and can be measured with simple computational basis measurements. The latter partition can be measured by the Bell basis measurement circuit in Figure~\ref{fig:bell_basis_measurement_circuit}.

\begin{figure}
    \centering
    {\includegraphics[width=.48\textwidth]{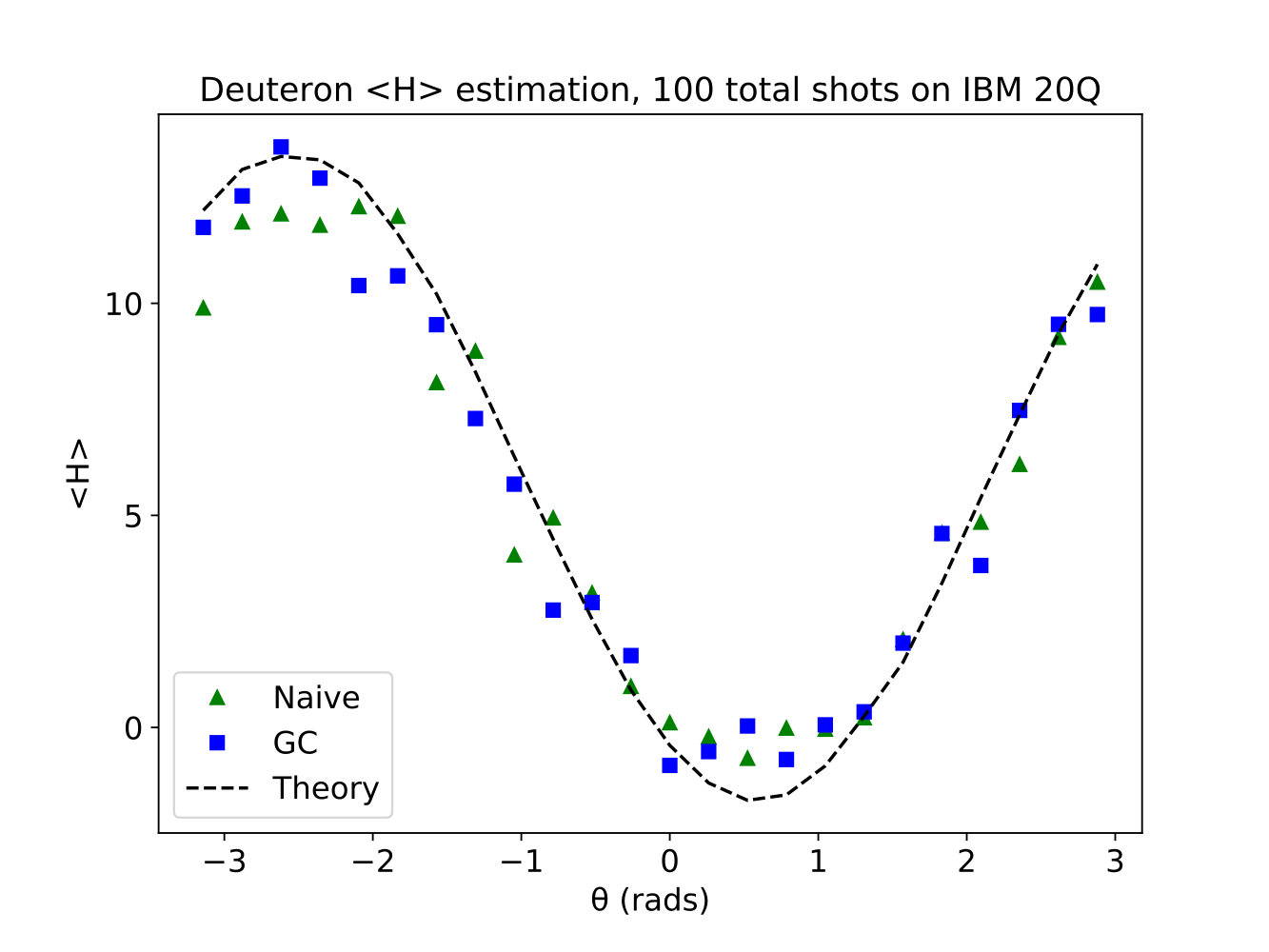}}
    {\includegraphics[width=.48\textwidth]{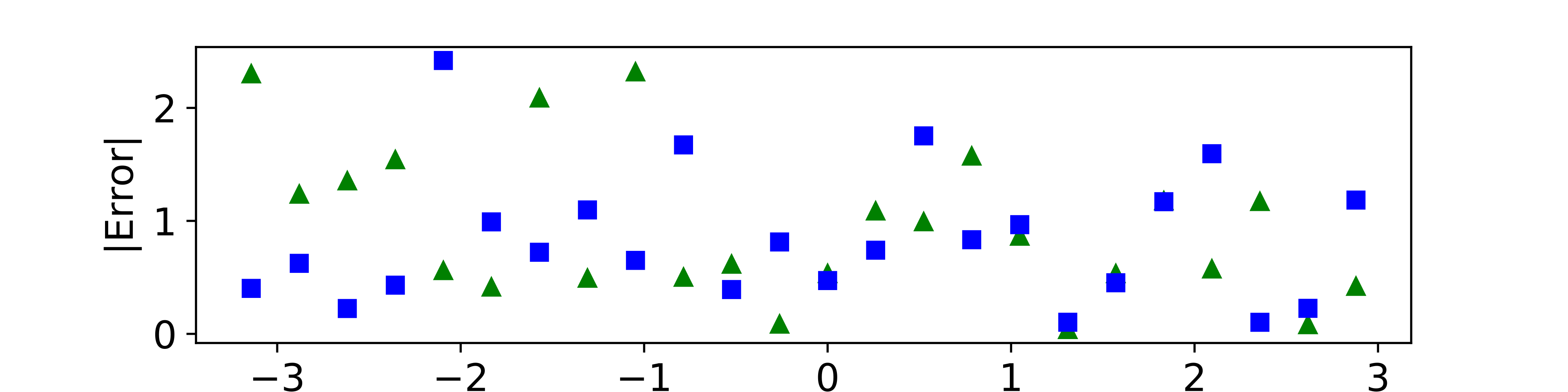}}
    \caption{Deuteron energy estimation under Naive and GC partitions, as executed on IBM Q20 with a total shot budget of 100. The energies are in MeV. Average error is 11\% lower with GC simultaneous measurement than with Naive separate measurements.}
    \label{fig:deuteron_100}
\end{figure}

To establish a fair comparison between Naive measurement and simultaneous measurement we performed experiments in which both settings were allocated an equal budget in total number of shots (trials) allowed. We first considered a resource-constrained setting with a budget of 100 total shots. This corresponds to 25 shots per partition in Naive measurement and 50 shots per partition in GC simultaneous measurement. Figure~\ref{fig:deuteron_100} plots our results for a simplified Unitary Coupled Cluster ansatz with a single parameter and just three gates (two single qubit rotations and one CNOT), as described in \cite{dumitrescu2018cloud}.

The results indicate reasonable agreement between Naive measurement, GC measurement, and the true (Theory) values. The deviation from Theory stems both from statistical variance due to the low shot budget, as well as systematic noise in the quantum processes. As Figure~\ref{fig:deuteron_100}'s lower |Error| plot indicates, for 13 of the 24 values swept across the $\theta$ range, GC measurement had lower error than Naive measurement. On average, the GC measurements had an error of 835 KeV---11\% less than the average error of 940 KeV for Naive measurement.

\begin{figure}
    \centering
    {\includegraphics[width=.48\textwidth]{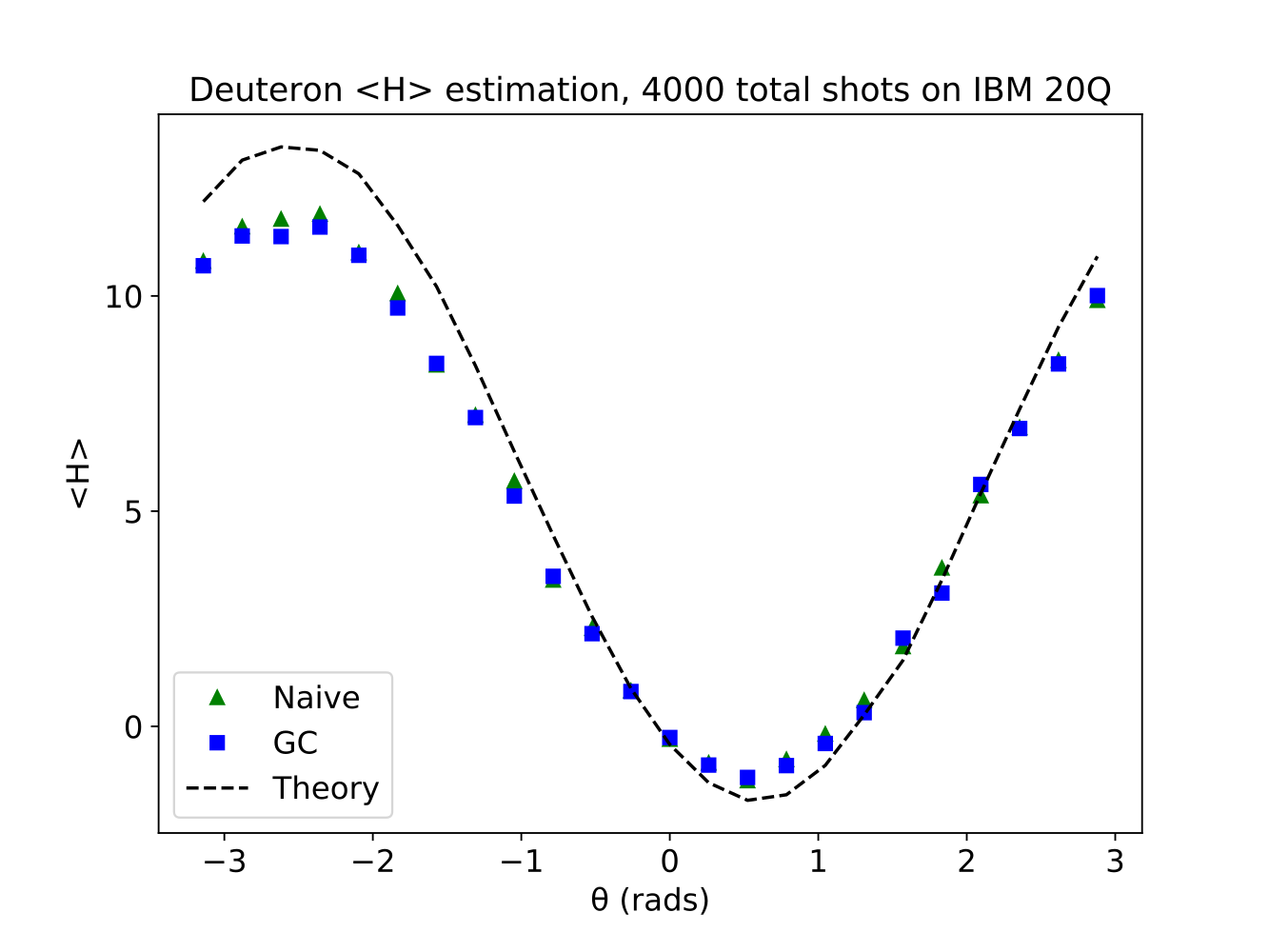}}
    {\includegraphics[width=.48\textwidth]{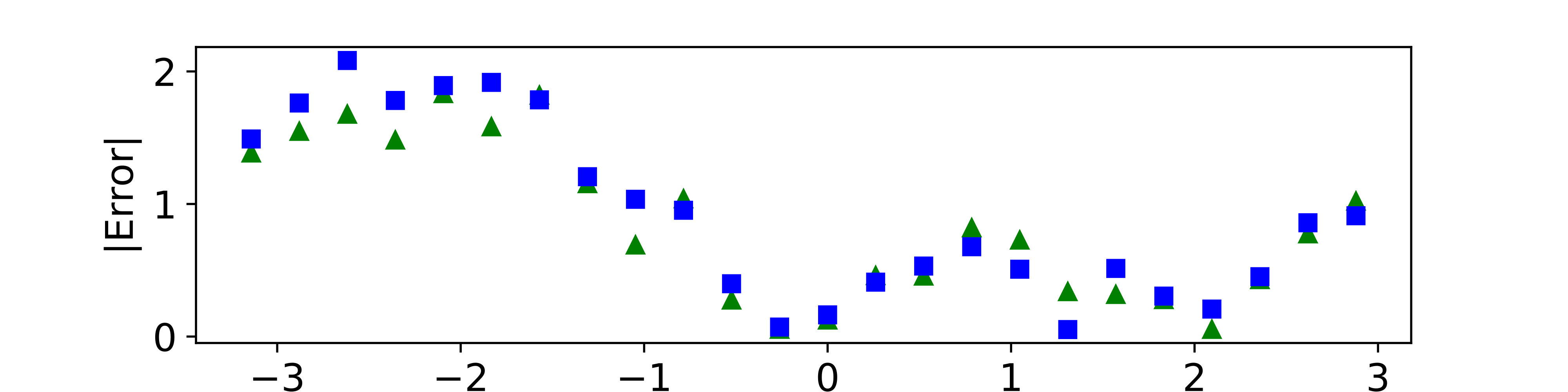}}
    \caption{Deuteron energy estimation under Naive and GC partitions, as executed on IBM Q20 with a total shot budget of 4000. The energies are in MeV. Average error is 7\% lower with Naive separate measurements than with GC simultaneous measurements.}
    \label{fig:deuteron_4000}
\end{figure}

We also ran another experiment with a much higher total shot budget of 4000 (i.e. 1000 shots per partition in Naive and 2000 for GC). In this regime, errors due to systematic quantum noise should dominate over errors from statistical variation. We expect GC simultaneous measurement to exhibit more systematic noise because it requires an extra CNOT gate as per the Bell measurement circuit in Figure~\ref{fig:bell_basis_measurement_circuit}. Therefore, we expect better results from Naive measurement than from GC simultaneous measurement.  Figure~\ref{fig:deuteron_4000} plots the experimental results.

For 17 of the 24 values swept across the $\theta$ range, Naive measurement does indeed outperform GC simultaneous measurement in terms of lower error. The respective average errors are 848 KeV and 914 KeV, indicating a 7\% higher accuracy with Naive measurement.

These results are presented as proof-of-concept that simultaneous measurement achieves higher accuracy when the shot budget is limited. Equivalently, we can achieve equal accuracy with fewer shots (i.e. fewer state preparations) when the shot budget is limited. For several reasons, we note that these experimental results \textit{underestimate} the potential of simultaneous measurement, especially as higher quantum volume devices emerge. In particular:
\begin{itemize}
    \item the Unitary Coupled Cluster ansatz of \cite{dumitrescu2018cloud} is highly simplified and does not yet exhibit the asymptotic $O(N^4)$ scaling. Our argument that simultaneous measurement is cheap hinges on the comparison between $O(N^4)$ ansatz gate count and $O(N^2)$ simultaneous measurement gate count. For this simplified ansatz and small $N$, simultaneous measurement essentially doubled the gate count. As lower-error devices emerge with the ability to support the full UCCSD ansatz gate count and larger qubit count $N$, simultaneous measurement circuits will become a negligible cost.
    \item For a small Hamiltonian like the one considered here, the partitioning gain from GC is only 2x. As indicated in the benchmark results in Section~\ref{sec:benchmark_results}, we expect up to 30x gains for larger Hamiltonians and possibly a gain factor that continues to linearly increase for larger molecules, based on extrapolation of the benchmark results.
    \item For current machines, the number of jobs is far more costly than the number of shots for practical purposes, since executions are scheduled at the granularity of jobs. In our executions, we saw this as an immediate and practical advantage of simultaneous measurement. Our total latency was dominated by the number of jobs rather than the number of shots, so our simultaneous measurement results were collected much more rapidly than Naive measurement results, even though both settings had equal total shot budgets.
\end{itemize}

We re-iterate that these results should only be interpreted as a proof of concept. As machines improve, we expect to see dramatically better results, for the aforementioned reasons.
\section{Statistics of Simultaneous Measurement: Guarding Against Covariances} \label{sec:covariance_reduction}
We have now shown both how to approximate a MIN-COMMUTING-PARTITION and how to actually construct the requisite simultaneous measurement circuits.  Finally, we now address an important question regarding the statistics of simultaneous measurement. This question was first raised by \cite[Section IV B2]{mcclean2016theory} which proved that simultaneous measurement can actually \textit{underperform} separate measurements due to the presence of covariance terms. In particular, while simultaneous measurement does not bias the estimate $\widehat{\braket{H}}$, it can increase the variance of the estimator, relative to separate measurements.

In this section, we first show a specific example from \cite{mcclean2016theory} in which simultaneous measurement is \textit{suboptimal}. Then, we prove that such examples are atypical and that the MIN-COMMUTING-PARTITION is still optimal when we have no prior on the ansatz state. Finally, we demonstrate an adaptive strategy for detecting and correcting course in the atypical case when a simultaneous measurement should be split into separate measurements.

\subsection{An Example} \label{subsec:an_example}
\begin{figure}[H]
\centering
\includegraphics[width=0.4\textwidth]{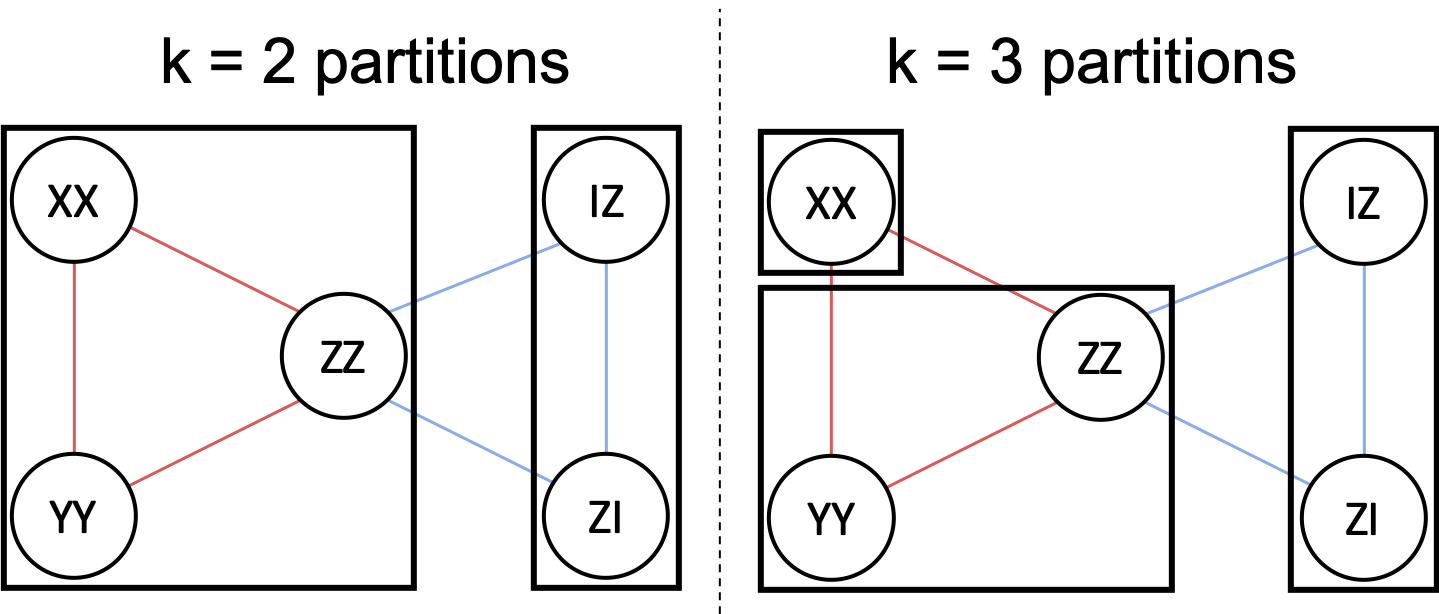}
\caption{Commuting-family partitions of $H = IZ + ZI - XX - YY + ZZ$ with $k = 2$ and $k = 3$.}
\label{fig:k2_and_k3}
\end{figure}

Consider the Hamiltonian, $H = IZ + ZI - XX - YY + ZZ$, following the example of \cite{mcclean2016theory}. The commutation graph has a bowtie shape. Figure~\ref{fig:k2_and_k3} depicts two possible clique partitionings with $k = 2$ and $k = 3$ commuting-family partitions respectively.

Thus far, we have worked under the assumption that estimating $\braket{H}$ is more efficient with simultaneous measurement than with separate measurements and we have therefore targeted MIN-COMMUTING-PARTITIONs. However, consider a case in which the ansatz state is $\ket{01}$, for the previously stated Hamiltonian. 

Since the outcomes of our measurements are random, we quantify the uncertainty around our estimate of the expectation value by $Var(\braket{H})$. Our end goal is to determine the expected value of the Hamiltonian to a target accuracy level $\epsilon$. 
The expected number of state preparations, $n_\text{expect}$, needed to achieve this accuracy for a $k$-way partitioning is \cite{mcclean2016theory}:
\begin{equation}
\label{eq:n_expect}
n_\text{expect} = \frac{k \sum_{i=1}^k Var(\text{Partition $i$})}{\epsilon^2}
\end{equation}

The variance from each partition can be computed from the formula for the variance of a sum of terms:
$$Var(\{\sum_{i=1}^n M_i\}) = \sum_{i=1}^n Var(M_i) + 2 \sum_{1 \leq i < j \leq n} Cov(M_i, M_j)$$
where $Cov(M_1, M_2) = \braket{M_1 M_2} - \braket{M_1} \braket{M_2}$ and $Var(M) = Cov(M, M)$.

In our case with $\ket{\psi} = \ket{01}$, the primitives evaluate to: $Var(IZ) = Var(ZI) = Var(ZZ) = 0$ and  $Var(-XX) = Var(-YY) = 1$. All covariances are 0 except for $Cov(-XX, -YY) = 1$.

For the $k = 2$ partitioning, we have
\begin{align*}
    n_{\text{expect}} =  \\
    \frac{2 \left[ Var(\{-XX, -YY, ZZ\}) + Var(\{ZI, IZ\}) \right] }{\epsilon^2} = \\
    2 \Big[ Var(-XX) + Var(-YY) + Var(ZZ) + \\ 2Cov(-XX, -YY) + 2Cov(-XX, ZZ) + 2Cov(-YY, ZZ) + \\
    Var(ZI) + Var(IZ) + 2Cov(IZ,ZI) \Big]  / \epsilon^2\\
    = \boxed{8 / \epsilon^2}
\end{align*}

For the $k = 3$ partitioning, we have:
\begin{align*}
    n_{\text{expect}} = \\
    \frac{3 \big[ Var(\{-XX\}) + Var(\{-YY, ZZ\}) + Var(\{IZ, ZI\}) \big] }{\epsilon^2} = \\
    = 3 \Big[ Var(-XX) + Var(-YY) + Var(ZZ)  + \\
    2Cov(-YY, ZZ) + Var(ZI) + Var(IZ) + 2Cov(IZ,ZI) \Big]  / \epsilon^2\\
    = \boxed{6 / \epsilon^2}
\end{align*}

Thus, due to the contribution of positive covariance between $-XX$ and $-YY$, the $k=3$ partitioning is better than the $k=2$ partitioning for this ($H, \ket{\psi}$) combination.

This phenomenon motivates us to pay close attention to covariances within each partitioning. The worst case scenario is that we end up with positive covariances within each partition. In a best case scenario, we'll have negative covariances within each partitioning, which could dramatically reduce the number of state preparations needed to achieve some desired error on $\braket{H}$.

\subsection{Typical Case} \label{subsec:typical_case}
We now observe that examples such as the previous one, in which the MIN-COMMUTING-PARTITION is suboptimal, are atypical. Below, we prove that when we have no prior on the ansatz state $\ket{\psi}$, the expected covariance between two commuting Pauli strings is 0. This validates the general goal of finding the MIN-COMMUTING-PARTITION, because under 0 covariances, the only strategy for reducing $n_{\text{expect}}$ in Equation~\ref{eq:n_expect} is to minimize the total number of partitions $k$.

\begin{theorem}
Given $M_1, M_2$, two commuting but non-identical Pauli strings, $\mathbb{E}[Cov(M_1, M_2)] = 0$ where the expectation is taken over a uniform distribution over all possible state vectors (the Haar distribution \cite{zyczkowski1998volume, russell2017direct}).
\end{theorem}
 
\begin{proof}
We consider the following two exhaustive cases:
\begin{enumerate}
    \item Either $M_1$ or $M_2$ is $I$. WLOG, suppose $M_1 = I$. Then, $Cov(M_1, M_2) = \braket{I \cdot M_2} - \braket{I} \braket{M_2} = 0$.
    \item Neither $M_1$ nor $M_2$ is $I$. Since $M_1$ and $M_2$ are Pauli strings which have only $+1$ and $-1$ eigenvalues, the eigenspace can be split into $M_1, M_2 = (-1, -1)$, $(-1, +1)$, $(+1, -1)$, and $(+1, +1)$ subspaces. Moreover, these subspaces are equally sized (proof follows from stabilizer formalism \cite[Chapter 10.5.1]{nielsen2010quantum}). Let us write $\ket{\psi}$ as a sum over projections into these subspaces:
    $$\ket{\psi} = a \ket{\psi_{-1,-1}} + b \ket{\psi_{-1,+1}} + c \ket{\psi_{+1,-1}} + d \ket{\psi_{+1,+1}}$$
    Under this state, the covariance is $Cov(M_1, M_2)_{\ket{\psi}} = \braket{M_1 M_2} - \braket{M_1} \braket{M_2} = (|a|^2 - |b|^2 - |c|^2 + |d|^2) - (-|a|^2 - |b|^2 + |c|^2 + |d|^2)(-|a|^2 + |b|^2 - |c|^2 + |d|^2)$.
    
    Now consider the matching state:
    $$\ket{\psi'} = b \ket{\psi_{-1,-1}} + a \ket{\psi_{-1,+1}} + d \ket{\psi_{+1,-1}} + c \ket{\psi_{+1,+1}}$$
    Under $\ket{\psi'}$, the covariance is
    $Cov(M_1, M_2)_{\ket{\psi'}} = \braket{M_1 M_2} - \braket{M_1} \braket{M_2} = (|b|^2 - |a|^2 - |d|^2 + |c|^2) - (-|b|^2 - |a|^2 + |d|^2 + |c|^2)(-|b|^2 + |a|^2 - |d|^2 + |c|^2)$.

    Thus, $Cov(M_1, M_2)_{\ket{\psi}} = - Cov(M_1, M_2)_{\ket{\psi'}}$. Since each $\ket{\psi}$ is matched by this symmetric $\ket{\psi'}$ state, and our expectation is over a uniform distribution of all possible state vectors, we conclude that $\mathbb{E}[Cov(M_1, M_2)] = 0$.
\end{enumerate}
\end{proof}

\subsection{Mitigating Covariances: Partition Splitting}

While we have now secured the top level goal of initially performing measurements under the MIN-COMMUTING-PARTITION approximation, it is still important to detect and correct course if covariances do turn out to harm our measurement statistics. We now introduce such a strategy that adaptively splits partitions to mitigate harmful covariances.

Our strategy is based on building sample covariance matrices of commuting Pauli strings. If  $M_1$, $M_2$, and $M_3$ are Pauli strings, recall that the covariance matrix, $Cov([M_1,M_2,M_3])$, under a fixed state is expressed as follows:
\[\left(\begin{array}{ccc}
Var(M_1) & Cov(M_1,M_2) & Cov(M_1,M_3)\\
Cov(M_2,M_1) & Var(M_2) & Cov(M_2,M_3)\\
Cov(M_3,M_1) & Cov(M_3,M_2) & Var(M_3)
\end{array}\right)
 \]
 Or, in shorthand notation, where $Var(M_1) = \sigma^2_{M_1}$ and $Cov(M_1,M_2) = \sigma_{M_1 M_2}$:
 \[
\left(\begin{array}{ccc}
\sigma^2_{M_1} & \sigma_{M_1 M_2} & \sigma_{M_1 M_3}\\
\sigma_{M_2 M_1} & \sigma^2_{M_2} & \sigma_{M_2 M_3}\\
\sigma_{M_3 M_1} & \sigma_{M_3 M_2} & \sigma^2_{M_3}\\
\end{array}\right)
 \]
 
Note that for commuting matrices $M_1$ and $M_2$, we have $Cov(M_1, M_2) = \braket{M_1 M_2} - \braket{M_1} \braket{M_2} = \braket{M_2 M_1} - \braket{M_2}\braket{M_1} = Cov(M_2, M_1)$, so covariance matrices are symmetric around the main diagonal.

We now return to the pathological example from Section~\ref{subsec:an_example}. Since the variance of a partitioning is the sum of all entries in each partition's covariance matrix, the sum of the shaded terms below represents the variance of the $k = 2$ partitioning ($\{-XX, -YY, ZZ\}, \{ZI, IZ\}$):
 
\[\left(\begin{array}{ccccc}
\shade \sigma^2_{-XX} & \shade \sigma_{-XX,-YY} & \shade \sigma_{-XX,ZZ} & \sigma_{-XX,ZI} & \sigma_{-XX,IZ} \\
\shade \sigma_{-YY,-XX} & \shade \sigma^2_{-YY} & \shade \sigma_{-YY,ZZ} & \sigma_{-YY,ZI} & \sigma_{-YY,IZ} \\
\shade \sigma_{ZZ,-XX} & \shade \sigma_{ZZ,-YY} & \shade \sigma^2_{ZZ} & \sigma_{ZZ,ZI} & \sigma_{ZZ,IZ} \\
\sigma_{ZI,-XX} & \sigma_{ZI,-YY} & \sigma_{ZI,ZZ} & \shade \sigma^2_{ZI} & \shade \sigma_{ZI,IZ} \\
\sigma_{IZ,-XX} & \sigma_{IZ,-YY} & \sigma_{IZ,ZZ} & \shade \sigma_{IZ,ZI} & \shade \sigma^2_{IZ} \\
\end{array}\right)
\]

And the sum of the shaded terms below represents  the variance of the $k=3$ partitioning ($\{-XX\}, \{-YY, ZZ\}, \{ZI, IZ\}$):

\[\left(\begin{array}{ccccc}
\shade \sigma_{-XX}^2 & \sigma_{-XX,-YY} & \sigma_{-XX,ZZ} & \sigma_{-XX,ZI} & \sigma_{-XX,IZ} \\
 \sigma_{-YY,-XX} & \shade \sigma_{-YY}^2 & \shade \sigma_{-YY,ZZ} & \sigma_{-YY,ZI} & \sigma_{-YY,IZ} \\
\sigma_{ZZ,-XX} & \shade \sigma_{ZZ,-YY} & \shade \sigma_{ZZ}^2 & \sigma_{ZZ,ZI} & \sigma_{ZZ,IZ} \\
\sigma_{ZI,-XX} & \sigma_{ZI,-YY} & \sigma_{ZI,ZZ} & \shade \sigma_{ZI}^2 & \shade \sigma_{ZI,IZ} \\
\sigma_{IZ,-XX} & \sigma_{IZ,-YY} & \sigma_{IZ,ZZ} & \shade \sigma_{IZ,ZI} & \shade \sigma_{IZ}^2 \\
\end{array}\right)
\]
\\

Therefore, it is favorable (fewer state preparations needed to achieve a target accuracy) to break the $-XX$ term out of the $\{-XX, -YY, ZZ\}$ partition if the condition atop the next page holds. The matrices represent a sum over enclosed terms, and the multiplicative factors of $k=2$ and $k=3$ follow from Equation~\ref{eq:n_expect}.
\pagebreak
\begin{widetext}
\[2 \left(\begin{array}{ccccc}
\shade \sigma_{-XX}^2 & \shade \sigma_{-XX,-YY} & \shade \sigma_{-XX,ZZ} & & \\
\shade \sigma_{-YY,-XX} & \shade \sigma_{-YY}^2 & \shade \sigma_{-YY,ZZ} & & \\
\shade \sigma_{ZZ,-XX} & \shade \sigma_{ZZ,-YY} & \shade \sigma_{ZZ}^2 & & \\
 & & & \shade \sigma_{ZI}^2 & \shade \sigma_{ZI,IZ} \\
 & & & \shade \sigma_{IZ,ZI} & \shade \sigma_{IZ}^2 \\
\end{array}\right)
 > 
3 \left(\begin{array}{ccccc}
\shade \sigma_{-XX}^2 & & & & \\
 & \shade \sigma_{-YY}^2 & \shade \sigma_{-YY,ZZ} & &  \\
 & \shade \sigma_{ZZ,-YY} & \shade \sigma_{ZZ}^2 & & \\
 & & & \shade \sigma_{ZI}^2 & \shade \sigma_{ZI,IZ} \\
 & & & \shade \sigma_{IZ,ZI} & \shade \sigma_{IZ}^2 \\
\end{array}\right)
\]
or equivalently, if:
\begin{equation} \label{eq:partition_splitting_covariance_matrix}
\left(\begin{array}{ccccc}
 & \shade 2 \sigma_{-XX,-YY} & \shade 2 \sigma_{-XX,ZZ} & & \\
\shade 2 \sigma_{-YY,-XX} & & & & \\
\shade 2 \sigma_{ZZ,-XX} & & & & \\
 & & & & \\
 & & & & \\
\end{array}\right)
 > 
\left(\begin{array}{ccccc}
\shade \sigma_{-XX}^2 & & & & \\
 & \shade \sigma_{-YY}^2 & \shade \sigma_{-YY,ZZ} & &  \\
 & \shade \sigma_{ZZ,-YY} & \shade \sigma_{ZZ}^2 & & \\
 & & & \shade \sigma_{ZI}^2 & \shade \sigma_{ZI,IZ} \\
 & & & \shade \sigma_{IZ,ZI} & \shade \sigma_{IZ}^2 \\
\end{array}\right)
\end{equation}

\end{widetext}

Informally, notice that the left-hand side of Equation \ref{eq:partition_splitting_covariance_matrix} is a multiple of the sum of the covariances that exist in the expression for $Var(k=2)$ but not $Var(k=3)$ (which we will call the ``broken terms"), whereas the right-hand side is a multiple of the sum of the variances and covariances that exist in both the $Var(k=2)$ and $Var(k=3)$ expressions (the ``unbroken terms"). This pattern generalizes such that it is favorable to switch from a partitioning with $k$ partitions to a clique-splitting partitioning with $k' > k$ partitions if:

\[k * (\sum \text{broken terms}) > (k' - k) * (\sum \text{unbroken terms})\]

A similar strategy was described in \cite[Section V. A.]{kandala2017hardware}, for the special case of comparing Naive partitions (with no covariances) with QWC partitions; our work generalizes to the case of comparing two non-Naive partitions where both sides have covariance terms.

\subsection{Strategies for covariance estimation}

As demonstrated in Section \ref{subsec:an_example}, the expected number of state preparations needed to determine $\braket{H}$ to an accuracy level $\epsilon$ can be calculated if the variances and pairwise covariances of commuting Pauli terms under an ansatz state are known.

In practice, the true theoretical values of these variances cannot be known beforehand, as doing so would require computations involving the exponentially sized ansatz state vector.  However, just as we use repeated measurements from partitions of commuting terms to approximate the expected value of their sum, \textbf{we can use these same measurements to approximate the covariance matrices of Pauli strings in the same partition}. This estimation of covariance is termed ``sample covariance", since its value is calculated via a sample from the theoretical distribution. This key idea of adaptively building a sample covariance matrix, using the measurements we are already making, allows us to adaptively detect and correct for harmful covariance terms.

Note that the theoretical variance of $\braket{M}$ is $Var(M) = \braket{M^2} - \braket{M}^2$, and is approximated by the sample variance, $\widehat{Var}(M) = \frac{1}{n-1} \sum_{i=1}^n (m_i - \overline {m})$, where $\{m_1,...,m_n\}$ represent the $n$ observed measurements of $M$, and where $\overline{m} = \frac{1}{n} \sum_{i=1}^n m_i$ is the sample mean.  Similarly, the theoretical covariance $Cov(M_1, M_2) = \braket{M_1 M_2} - \braket{M_1} \braket{M_2}$ is approximated by the sample covariance $\widehat{Cov}(M_1,M_2) = \frac{1}{n-1} \sum_{i=1}^n (m_{1i} - \overline {m_1}) (m_{2i} - \overline{m_2})$ where $\{m_{11},...,m_{1n}\}$ and $\{m_{21},...,m_{2n}\}$ are the $n$ observed measurements of $M_1$ and $M_2$ respectively.

Since covariance terms can only be approximated if terms are simultaneously measured, we ideally want to start our measurements in a setting with MIN-COMMUTING-PARTITIONS. Fortunately, this is exactly the optimal starting strategy that we initialize with, as per the argument in Section~\ref{subsec:typical_case}. Once we collect sufficiently many observations that the sample covariance matrices stabilize, this will enable us to identify opportunities to split partitions in order to lower variances and thus reduce the number of requisite state preparations.

To make this concrete, let us again consider the $k=2$ partitioning from the previous example, $\{-XX, -YY, ZZ\}, \{ZI, IZ\}$.  As we accumulate more observations, we can empirically build up an approximation of each partition's sample covariance matrices, like so:

 \[\left(\begin{array}{ccccc}
 \hat{\sigma}_{-XX}^2 & \hat{\sigma}_{-XX,-YY} & \hat{\sigma}_{-XX,ZZ} & & \\
\hat{\sigma}_{-YY,-XX} &  \hat{\sigma}_{-YY}^2 &  \hat{\sigma}_{-YY,ZZ} & & \\
\hat{\sigma}_{ZZ,-XX} &  \hat{\sigma}_{ZZ,-YY} &  \hat{\sigma}_{ZZ}^2 & & \\
& & & \hat{\sigma}_{ZI}^2 & \hat{\sigma}_{ZI,IZ}\\
& & & \hat{\sigma}_{IZ,ZI} & \hat{\sigma}_{IZ}^2 \\
\end{array}\right)
 \]
 
 Since the sample covariance matrix $\widehat{Var}(k=2)$ contains a superset of the terms needed to calculate $\widehat{Var}(k=3)$, we can use observations from the $k=2$ setting to explore whether further partitions would be beneficial.
 
\begin{figure}[h]
    \centering
    {\includegraphics[width=.48\textwidth]{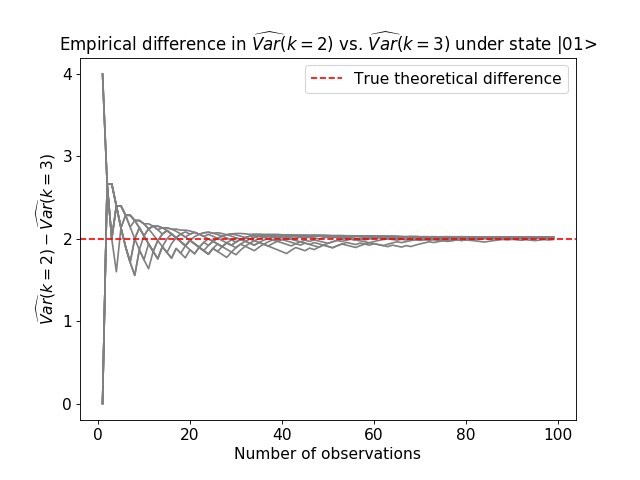}}
    \caption{Convergence of the empirical difference, $\widehat{Var}(k=2) - \widehat{Var}(k=3)$, to the true difference in variances under $\ket{01}$.  Since $Var(k=2) - Var(k=3)$ is positive, this signals that the $k=3$ partitioning will lead to a lower-variance estimator.}
    \label{fig:empirical_covariance_state01}
\end{figure}

 Each of the grey lines in Figure \ref{fig:empirical_covariance_state01} depicts the value of $\widehat{Var}(k=2) - \widehat{Var}(k=3)$ as it evolves with a set of 100 observed measurements under the $\ket{01}$ state.  The plot illustrates that the empirical difference, $\widehat{Var}(k=2) - \widehat{Var}(k=3)$ converges to the true theoretical difference, $Var(k=2) - Var(k=3)=2$ after around 30 observations. The positive sign of this difference indicates that $Var(k=3) < Var(k=2)$, and therefore the $k=3$ partitioning should be favored due to its lower variance.
 
\begin{figure}[h]
    \centering
    {\includegraphics[width=.48\textwidth]{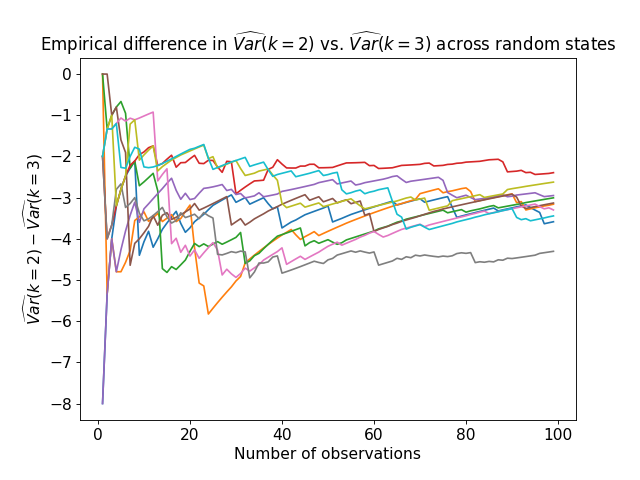}}
    \caption{The empirical difference in $\widehat{Var}(k=2) - \widehat{Var}(k=3)$ across ten Haar-randomly-chosen states. While the convergence value differs across states, it is is negative in all ten cases. This contrasts with the atypical case of convergence to a positive value in the example of Figure \ref{fig:empirical_covariance_state01} under state $\ket{01}$.}
    \label{fig:empirical_covariance_randomstates}
\end{figure}

When we broaden analysis of the $k=2$ versus $k=3$ setting across many different random states, we observe that the state $\ket{01}$ is indeed atypical and pathological, as suggested in Section~\ref{subsec:typical_case}. Under the vast majority of states, the variance of the $k=2$ setting is lower than the $k=3$ setting, as observed by the negative values of $\widehat{Var}(k=2) - \widehat{Var}(k=3)$ in Figure \ref{fig:empirical_covariance_randomstates}, and therefore the $-XX$ term should not be split into a separate partition.

This discussion naturally leads to the question of how many observations are necessary for the sample covariance matrix to be a good approximation of the true theoretical covariance matrix.  To answer this question, we need to formalize a notion of the accuracy of a sample covariance matrix.  Several candidate measures may be considered, which we are exploring in ongoing work:
\begin{itemize}
    \item Enforcing a minimum number of ``burn-in" observations.  This acts as a proxy of the sample observations being sufficiently representative of the true theoretical distribution.
    \item Enforcing that the distance between the sample covariance matrix after $n-1$ observations and after $n$ observations be less than a pre-specified threshold.  This acts as an alternative proxy of the stability of the observations on which the sample covariance matrix is based.
    \item Enforcing that a hypothesis test between the sample variance of the full partitioning and the sample variance of the split-up partitioning returns a p-value below a pre-specified significance level.
\end{itemize}
    
The last candidate measure is the most attractive because p-values can be compared across different experimental settings.  By contrast, appropriate cutoff values for the first two measures vary with $H$ and $\ket{\psi}$.  Formalization of the last measure will require further work to confirm the distribution of the sample variance and covariance terms.
\section{Conclusion} \label{sec:conclusion}

Our techniques and demonstrations show that simultaneous measurement substantially reduces the cost of Variational Quantum Eigensolver by allowing state preparations to cover several Pauli strings simultaneously. We demonstrate algorithms that achieve up to 30x reductions in the number of requisite state preparations. We also raise practical concerns about these algorithms and identify an alternate strategy that exploits properties of molecular Hamiltonians to achieve an 8x reduction in state preparation cost, with almost no additional pre-computation. Our systems emphasis includes explicit attention to the overhead of simultaneous measurement circuits. Accordingly, we develop a circuit synthesis procedure, which we have implemented and tested in software. We also study the statistics of simultaneous measurement, and ensure that the top-level goal of finding MIN-COMMUTING-PARTITIONs is statistically justified. Our statistical analysis also yields a strategy for detecting and correcting course when simultaneous measurements are harmed by covariance terms. Our theoretical and benchmark/simulation results are accompanied by a proof-of-concept experimental validation on the IBM 20Q quantum computer.

Our ongoing work includes further benchmarking, more theoretical investigation, and the development of a software tool that packages together all of our techniques. We also see promising future work towards further developing molecular-Hamiltonian-aware partitioning strategies, especially since the advantage of the MIN-COMMUTING-PARTITION appears to improve with molecular size. Moreover, other qubit encodings like Bravyi-Kitaev, as well as Hamiltonian reduction techniques such as active space reductions and frozen orbitals should be considered.




\appendix
\appendix
\section{MIN-COMMUTING-PARTITION is NP-Hard}
\label{app:np_hard}
\begin{figure}[h]
    \centering
    \includegraphics[width=0.45\textwidth]{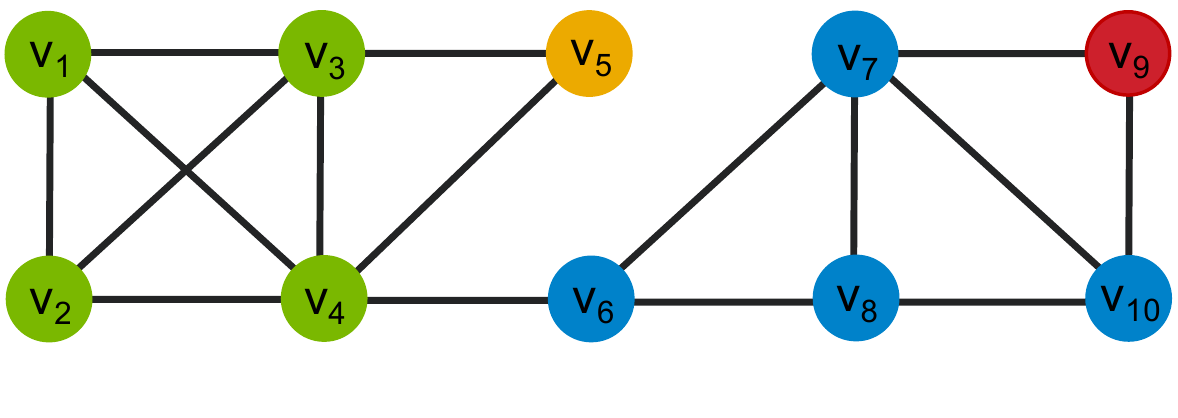}
    \begin{blockarray}{ccccccccccc}
            & $v_1$ & $v_2$ & $v_3$ & $v_4$ & $v_5$ & $v_6$ & $v_7$ & $v_8$ & $v_9$ & $v_{10}$\\
    $o_1$:    & Z & I & I & I & X & X & X & X & X & X \\
    $o_2$:    & I & Z & I & I & X & X & X & X & X & X \\
    $o_3$:    & I & I & Z & I & I & X & X & X & X & X \\
    $o_4$:    & I & I & I & Z & I & I & X & X & X & X \\
    $o_5$:    & I & I & I & I & Z & X & X & X & X & X \\
    $o_6$:    & I & I & I & I & I & Z & I & I & X & X \\
    $o_7$:    & I & I & I & I & I & I & Z & I & X & I \\
    $o_8$:    & I & I & I & I & I & I & I & Z & X & I \\
    $o_9$:    & I & I & I & I & I & I & I & I & Z & I \\
    $o_{10}$: & I & I & I & I & I & I & I & I & I & Z \\
    \end{blockarray}
\caption{Instance of MIN-CLIQUE-COVER (top) and MIN-COMMUTING-PARTITION (bottom).}
\label{fig:reduction}
\end{figure}

We show that MIN-COMMUTING-PARTITION is NP hard. Given a set of operators $o_1, o_2, …, o_n$, the MIN-COMMUTING-PARTITION problem partitions the operator set into $k$ subsets such that all operators in each subset pairwise commute and $k$ is minimized. The corresponding decision problem is in NP as it is easy to verify pairwise commutativity for each subset of operators. To show NP completeness it remains to show the problem is NP hard. This can be done by reducing from MIN-CLIQUE-COVER. Given a graph $G=(V,E)$ with $n$ vertices that represents an instance of MIN-CLIQUE-COVER, we produce an instance of MIN-COMMUTING-PARTITION consisting of a set of operators $o_1, o_2, …, o_n$ where each operator $o_i$ has $n$ Paulis, and the j-th Pauli is $Z$ if $j=i$, $X$ if $j>i$ and $(v_i,v_j) \not\in E$, and $I$ otherwise. This is illustrated in Figure~\ref{fig:reduction}. It is easy to see that a commuting subset of operators corresponds to a clique, which concludes the proof. Notice that the commutativity relationships required in this reduction are only Qubit-Wise Commutative, meaning that even the QWC-restricted MIN-COMMUTING-PARTITION problem is NP-Hard.

\section*{Acknowledgements}
We would like to thank Kenneth Brown, Peter Love, and Will Kirby for helpful discussions about VQE. We are also grateful to Olivia Di Matteo for introducing us to MUBs and clarifying the construction of simultaneous measurement circuits. Additionally, we acknowledge Henry Hoffmann for advising us on the time-to-solution for classical optimizers.

This work is funded in part by EPiQC, an NSF Expedition in Computing, under grants CCF-1730449/1730082, and in part by STAQ, under grant NSF Phy-1818914. P. G. is supported by the Department of Defense (DoD) through the National Defense Science \& Engineering Graduate Fellowship (NDSEG) Program. O. A. is supported by the National Science Foundation Graduate Research Fellowship Program under Grant No. DGE 1752814. The work of K. G. and M. S. is supported by the U.S. Department of Energy, Office of Science, under contract number DE-AC02-06CH11357.

\bibliographystyle{unsrt}
\bibliography{references}

\end{document}